\documentclass[twocolumn]{aastex63}

\usepackage{booktabs}
\usepackage{amsmath}
\usepackage{soul}
\usepackage{lineno}
\linenumbers

\newcommand{\be}{\begin{equation}}
\newcommand{\ee}{\end{equation}}

\newcommand\appendixfootnote[1]{\footnote{\hsize=\columnwidth\advance\hsize by-18pt\relax#1}}

\received{}
\revised{}
\accepted{}
\submitjournal{}

\begin{document}

\title{Constraining scalar-tensor theories by  neutron star-balck hole gravitational wave events}

\correspondingauthor{Rui Niu, Wen Zhao}
\email{nrui@mail.ustc.edu.cn, wzhao7@ustc.edu.cn}

\author[0000-0001-9098-6800]{Rui Niu}
\affiliation{CAS Key Laboratory for Researches in Galaxies and Cosmology, Department of Astronomy, University of Science and Technology of China, Chinese Academy of Sciences, Hefei, Anhui 230026, China;}
\affiliation{School of Astronomy and Space Sciences, University of Science and Technology of China, Hefei 230026, China}


\author[0000-0001-5435-6502]{Xing Zhang}
\affiliation{CAS Key Laboratory for Researches in Galaxies and Cosmology, Department of Astronomy, University of Science and Technology of China, Chinese Academy of Sciences, Hefei, Anhui 230026, China;}
\affiliation{School of Astronomy and Space Sciences, University of Science and Technology of China, Hefei 230026, China}

\author[0000-0002-3784-8684]{Bo Wang}
\affiliation{CAS Key Laboratory for Researches in Galaxies and Cosmology, Department of Astronomy, University of Science and Technology of China, Chinese Academy of Sciences, Hefei, Anhui 230026, China;}
\affiliation{School of Astronomy and Space Sciences, University of Science and Technology of China, Hefei 230026, China}

\author[0000-0002-1330-2329]{Wen Zhao}
\affiliation{CAS Key Laboratory for Researches in Galaxies and Cosmology, Department of Astronomy, University of Science and Technology of China, Chinese Academy of Sciences, Hefei, Anhui 230026, China;}
\affiliation{School of Astronomy and Space Sciences, University of Science and Technology of China, Hefei 230026, China}

\begin{abstract}

With the continuous upgrade of detectors, more and more gravitational wave (GW) events were captured by the LIGO Scientific Collaboration and Virgo Collaboration (LVC), which offers a new avenue to test General Relativity and explore the nature of gravity. 
Although, various model-independent tests have been performed by LVC in previous works, it is still interesting to ask what constraints on specific models can be placed by current GW observations.  
In this work, we focus on three models of scalar-tensor theories, the Brans-Dicke theory (BD), the theory with scalarization phenomena proposed by Damour and Esposito-Far\`{e}se (DEF), and Screened Modified Gravity (SMG). 
{From all 4 possible NSBH events so far, we use two of them to place the constraints. The other two are excluded in this work due to the possible unphysical deviations. }
We consider the inspiral range with the cutoff frequency at the innermost stable circular orbit and add a modification of dipole radiation into the waveform template. The scalar charges of neutron stars in the dipole term are derived by solving the Tolman-Oppenheimer-Volkoff equations for different equations-of-states.  The constraints are obtained by performing the full Bayesian inference with the help of the open source software \texttt{Bilby}. 
The results show that the constraints given by GWs are comparable with those given by pulsar timing experiments for DEF theory, but are not competitive with the current solar system constraints for BD and SMG theories.
        
\end{abstract}

\keywords{gravitational wave, scalar-tensor theory}

\section{Introduction} \label{sec_intro}
The theory of General Relativity (GR), as one of the two pillars of modern physics, is regarded as the most beautiful theory by common consent \citep{Chandrasekhar1984}. 
The splendor of GR is not only due to its elegant mathematical expression, but also its precise consistency with experimental tests.
Since Einstein proposed GR in 1915, a large number of experimental tests have been conducted, ranging from laboratory scale \citep{Sabulsky2019,Hoyle2001,Adelberger2001}
to solar system scale \citep{Will2018,Will2014}
and to cosmological scale
\citep{Jain2010,Koyama2016,Clifton2012}. 
In recent years, pulsar timing experiments \citep{Stairs2003,Manchester2015,Wex2014,Kramer2017} and gravitational wave observations \citep{Abbott2019,Collaboration2020,Abbott2016a,Abbott2019c} provide great opportunities to test GR under strong field conditions.
So far, all these experimental tests have supported GR at a very high level of accuracy.

Although great success has been achieved, there are still problems that GR cannot solve.
At the theoretical level, GR has been facing difficulties such as the singularity and quantization problems \citep{DeWitt1967,Kiefer2007}. 
At the experimental level, to explain astrophysical and cosmological observation data within the GR framework, 
it is necessary to introduce the so-called dark matter and dark energy whose physical nature is still unknown, 
which might imply the incompleteness  of GR \citep{Cline2013,Sahni2004}.
With the motivation to solve these problems, many modified gravity theories have been proposed. 
Among them, the scalar-tensor theories are generally considered as a promising candidate \citep{YasunoriFujii2016}. 

The origin of scalar-tensor theories can be traced back to the works of Kaluza and Klein \citep{Kaluza1921,Klein1926}. The form that we are familiar with today was developed by works \citep{Brans1961,Fierz1956,Jordan1955}. The scalar-tensor theories have potential relations with dark energy, dark matter and inflation, which continually arouse people’s interest in contemporary \citep{Clifton2012,Barrow1990,Burd1991,Schimd2005,Kainulainen2006,Brax2006}.
We focus on three different models of scalar-tensor theories in this work, i.e., the Brans-Dicke theory (BD), the theory with scalarization phenomena proposed by Damour and Esposito-Far\`{e}se (DEF), and the screened modified gravity (SMG).
The theory of Brans and Dicke \citep{Brans1961} takes Mach's principle as the starting point, which tells the phenomenon of inertia depends on the mass distribution of the universe. Thus the gravitational constant is promoted to be variable and coupled to the Einstein-Hilbert Lagrangian as a scalar field \footnote{In practice, the possible dependence of $G$ on different circumstance is testable in cosmological scale \citep{Zhao2018}.}.
The Brans-Dicke theory is the simplest scalar-tensor theory and is usually seen as the prototype of scalar-tensor theories, which has been well studied and constrained \citep{Will2018}.
Extensive tests have been performed in weak field regimes based on parameterized post-Newtonian formalism. The most stringent constrain is given by the measurement of Shapiro time delay from Cassini-Huygens spacecraft \citep{Bertotti2003}.

For Brans-Dicke theory, this tight bound requires deviations from GR in all gravitational experiments to be very small in both weak-field and strong-field. 
However, in the works of Damour and Esposito-Far\`{e}se \citep{Damour1993,Damour1992}, they showed that some nonperturbative effects can emerge in strong-field conditions. When the compactness of an object exceeds a critical point, a phenomenon, so-called spontaneous scalarization, which is usually discussed by analogy with the spontaneous magnetization in ferromagnets \citep{Damour1996}, will arise.
This phenomenon can make the behavior in gravitational experiments involving compact objects, like binary neutron star systems, have remarkable differences from the experiments in weak-field regimes. 
In the models that can develop nonperturbative strong-field effects, order-of-unity deviations from GR are still allowed in strong-field experiments, under the premise of passing the most stringent weak-field constraint.
In the subsequent researches, different kinds of scalarization phenomena, dynamical scalarization and induced scalarization, are discovered in numerical relativity simulations of merging binary neutron stars evolution \citep{Barausse2013}. 
In binary neutron star systems, the phenomenon that the scalar field produced by the scalarized component can induce the scalarization of another component which is not scalarized initially is called induced scalarization. Since the GW event used in this work is considered as a neutron star-black hole (NSBH) binary event, this phenomenon is not needed to be concerned. Dynamical scalarization is a phenomenon that a binary system, in which both two components cannot be scalarized in isolation, is triggered to scalarization due to their gravitational binding energy of orbit. 
However, in the previous works \citep{Palenzuela2014,Sampson2014}, it has been shown that dynamical scalarization is difficult to be detected by current detectors. Therefore, we only consider spontaneous scalarization in this work. 

The nonperturbative strong-field effects can be constrained by pulsar timing experiments \citep{Damour1998,Damour1996}. Because of precise measurement technology and decades of data accumulation, the orbital period decay rate of binary pulsar systems can be measured in high precision, which makes pulsar timing experiments a good tool to test gravitational theories in strong-field regimes \citep{Wex2014}. In previous works, stringent limits have been placed by using recent observational results from binary pulsar systems \citep{Zhao2019,Shao2017,Freire2012,Antoniadis2013,Cognard2017,Anderson2019}. 

There is another class of models, screened modified gravity (SMG), which can evade the tight solar system constraints by introducing screening mechanisms \citep{Clifton2012}. 
Various kinds of screen mechanisms have been introduced and studied, such as Chameleon mechanism \citep{Khoury2004,Khoury2004a}, Vainshtein mechanism \citep{Vainshtein1972,Babichev2013} and symmetron mechanism \citep{Hinterbichler2010}. 
The scalar field can be used to play the role of dark energy for driving the acceleration of the cosmic expansion in cosmological scales. Meanwhile, screening mechanisms can suppress deviations from GR in small scales to circumvent stringent constraints from the solar system tests and laboratorial experiments. (see \citep{Joyce2015,Clifton2012,Khoury2010,Brax2012} for comprehensive reviews).
Numerous tests on SMG have also been performed in different systems \citep{Burrage2018,Sakstein2020,Ishak2018,Zhang2019,Brax2014,Liu2018a,Zhang2019a,Niu2020,Zhang2019b}.

In recently, the first gravitational wave (GW) event GW150914 was directly detected by LIGO, which confirmed the last remaining not directly detected prediction of GR \citep{Abbott2016}. And more GW events are captured in the subsequent observing runs by the LIGO-Virgo collaborations (LVC) \citep{Abbott2019,Abbott2020}. 
With the continuing upgrades of sensitivity and the joining of new detectors, GW detections are becoming routines. The GW observations offer a new avenue to test GR and explore the nature of gravity in the extremely strong field regime.

LVC has performed various model-independent tests on observed events, and no evidence for deviations from GR has been found \citep{Abbott2019,Collaboration2020,Abbott2016a,Abbott2019c}. However, for a given specific modified gravity, the model-independent parameters always cannot completely describe the deviations of GWs, which naturally depend on the characters of neutron stars and/or black holes in the corresponding theory. Therefore, it is still interesting to see what constraints on specific models can be given by current observation, which are complementary with the model-independent tests. 
In this work, we consider three specific scalar-tensor theories mentioned above, BD, DEF and SMG.

Testing scalar-tensor theories by GW has been concerned since the 1990s \citep{Will1994}. Now, more and more detections of GW event and open access data allow us to constrain scalar-tensor theories by real GW data. Since in scalar-tensor gravities, the deviation of GW from that in GR depends on the sensitivity difference of two stars, the asymmetric binaries (e.g. NSBH, white dwarf-NS, white dwarf-BH binaries) are the excellent targets for the model tests.

{ So far, among all GW events captured by LVC, there are four possible NSBH events, GW200105, GW200115, GW190426\_152155 and GW190814 \citep{Abbott2021a,Abbott2020,Abbott2020a}. 
The two events, GW200105 and GW200115, released recently, are the first confident observations of NSBH binaries \citep{Abbott2021a}. The component masses of these two events are consistent with current observations of black holes and neutron stars.
However, the data are uninformative about the spin or tidal deformation, and no electromagnetic counterparts are detected. There is no direct evidence that the secondaries of these two events are neutron stars. Although it cannot be ruled out that the secondaries are some kind of exotic objects, we follow the most natural interpretation of these two events that they are NSBH coalescence events.

There are also two plausible NSBH events, GW190814 and GW190426\_152155 in the second Gravitational-Wave Transient Catalog (GWTC-2). But the nature of these two events is not definitively clear. 
The secondary mass of GW190814 is about $2.6M_\odot$, which could be interpreted as either a low-mass black hole or a heavy neutron star \citep{Abbott2020a, Most2020, Broadhurst2020}. However, according to current knowledge and observations of neutron stars, its lighter object is likely too heavy to be a neutron star \citep{Abbott2020a}. We exclude this event in our analysis.
Meanwhile, the event GW190426\_152155 has the highest false alarm rate (FAR) \citep{Abbott2020}. Whether it is a real signal of astrophysical origin is still not definitively clear yet. But its component masses are consistent with current understanding of black holes and neutron stars. There are many recent works concerning this event, such as \citep{Broekgaarden2021,Li2020,RomanGarza2020}. following some of them, we make our discussion on the assumption that the GW190426\_152155 is an NSBH coalescence event.
It needs to be emphasized that our analysis will be not applicable if this event is not a real NSBH binary.

There is another obstacle in our analysis. For events with a large mass ratio, deviations have been seen in the posterior distributions of the dipole modification parameter, in which the GR value is excluded from $90\%$ confidence intervals. 
The case of GW190814 has been shown in the previous works (referring to Appendix C in \citep{Collaboration2020} and Appendix A in \citep{Perkins2021} for more detail).
We have also seen similar deviations in our analysis of GW200105.
The deviations are believed to be unphysical effects which are probably caused by waveform systematics, covariances between parameters, or the way of non-GR modification parameterization.
To thoroughly explain these deviations, more studies about the parameterized tests of GR on highly asymmetric sources are needed. 
In this work, we exclude the event GW200105, and only employ the data of GW200115 and GW190426\_152155.}

The previous work \citep{Zhao2019} has used the binary neutron star event in GWTC-1, GW170817, to constrain scalarization effects. However, instead of directly using strain data, They employed the measurement of mass and radius from \citep{Abbott2018,Abbott2019a} to get the constraints. In this work, we use the modification of dipole radiation in waveform and perform the full Bayesian inference to constrain scalarization effects.

The rest of this paper is organized as follows. In the next section, the modified gravity models considered in this work, including BD, DEF and SMG, are briefly reviewed. 
Then, in Section \ref{sec_3}, we present the basic information and principle of data and statistical method used in this work.
The results and conclusions are discussed in Section \ref{sec_4}.
The formulae used to get scalar charges of neutron stars by solving Tolman-Oppenheimer-Volkoff (TOV) equations are presented in Appendix \ref{appendix-1} for convenience of reference. 
In Appendix \ref{appendix-2} and \ref{appendix-3}, we illustrate the comparisons of posterior distributions of other parameters with the posterior data released by LVC, and compare the constraints on the dipole radiation with the results reported by LVC.
We also present the scalar charges gotten from solutions of TOV equations for all four equations-of-state (EoS) considered in this work in Appendix \ref{appendix-4}.
{ A discussion of the other two possible NSBH events which are excluded in the work, GW190814 and GW200105, is presented in Appendix \ref{appendix-5}.}
All parameter estimation samples of this work are available on Zenodo\footnote{\url{https://doi.org/10.5281/zenodo.5188445}}. 
Throughout this paper, we use the units in which $\hbar=c=1$.

\section{Scalar-Tensor Theories} \label{sec_2}

In this work we consider a class of scalar-tensor theories, which can be described by the action
\begin{equation} \label{action}
\begin{aligned}
    S = & \frac{1}{16\pi G_*} \int{\rm d}^4 x \sqrt{-g_*} 
    \Bigl[R_* - 2g_*^{\mu\nu} \partial_\mu\varphi \partial_\nu\varphi \Bigr] \\
    & + S_m\Bigl[\psi_m, A^2(\varphi) g_*^{\mu\nu}\Bigr]
\end{aligned}
\end{equation}
in the Einstein-frame.
$G_*$ denotes the bare gravitational coupling constant, which is approximated by $G$ when solving TOV equations in practical. $g^{\mu\nu}_*$ and $g_*$ are the Einstein-frame metric and its determinant, and $R_*\equiv g^{\mu\nu}_*R^*_{\mu\nu}$ is the Ricci scalar.
The last term is the action of matter, where $\psi_m$ collectively denotes various matter fields and $A(\varphi)$ is the conformal coupling function.
Since the potential $V(\varphi)$ will be considered only in the SMG theory, we do not write the $V(\varphi)$ in the above action.
The field equations can be derived by varying the action (\ref{action}) with respect to the metric $g^{\mu\nu}_*$ and scalar field $\varphi$,
\begin{equation}\label{field_eqs}
\begin{aligned}
R^*_{\mu\nu} &= 2\partial_\mu\varphi \partial_\nu\varphi + 8\pi G_*\left(T^*_{\mu\nu} - \frac{1}{2} T^*g^*_{\mu\nu} \right), \\
\Box_{g_*}\varphi  &= -4\pi G_*\alpha(\varphi)T_*,
\end{aligned}
\end{equation}
where $\Box_{g_*} \equiv (-g_*)^{-1/2}\partial_\mu(\sqrt{-g_*}g^{\mu\nu}\partial_\nu)$ is the curved space D'Alembertian, $T_*^{\mu\nu}\equiv 2(-g_*)^{-1/2}\delta S_m/\delta g^*_{\mu\nu}$ is the energy-momentum tensor of matter fields and $T_*\equiv g_{\mu\nu}^* T_*^{\mu\nu}$.
The quantity $\alpha(\varphi)$ is defined as $\alpha(\varphi)\equiv \partial\ln A(\varphi)/\partial\varphi$, which describes the coupling strength between the scalar field and matters.
The $\ln A(\varphi)$ can be expanded around the background value $\varphi_0$ of the scalar field as
\be \label{expansion_of_coupling_func}
\ln A(\varphi) = \alpha_0(\varphi-\varphi_0) + \frac12\beta_0(\varphi-\varphi_0)^2 + \mathcal{O}(\varphi-\varphi_0)^3,
\ee
where the coefficients $\alpha_0$ and $\beta_0$ are related to two parameters $\beta^{\rm PPN}$ and $\gamma^{\rm PPN}$ in parameterized post-Newtonian (PPN) formalism by \citep{Will2018}
\begin{align}
\gamma^{\rm PPN} - 1 &= -\frac{2\alpha_0^2}{1+\alpha_0^2}, \\
\beta^{\rm PPN} -1 &= \frac12\frac{\alpha_0^2\beta_0}{(1+\alpha_0^2)^2}.
\end{align}

In the context of compact binary systems, a parameter called the scalar charge, which is defined as 
\be
\alpha_A \equiv \frac{\partial \ln m_A}{\partial \varphi} \bigg|_{\varphi=\varphi_0},
\ee
can describe the coupling between the scalar field and the star $A$. 
This parameter is used to determine the equation of motion and gravitational wave emission of binary systems.
For compact binaries in scalar-tensor theories, the center of gravitational binding energy and the center of inertial mass are not coincident, which results in the varying dipole moment and induces extra energy loss by dipole radiation \citep{Will1994}. 
We consider a gravitational waveform with the leading order of the modification, which has a dipole term in the phase \citep{Will1994,Tahura2018,Zhang2017a,Liu2018,Liu2020}, 
\be \label{dipole_term_1}
h(f) = h_{\rm GR}(f)\exp\left[ i \frac{3}{128\eta} \varphi_{-2} (\pi G M f)^{-7/3} \right], 
\ee
where $\varphi_{-2}$ is given by 
\be \label{dipole_term_2}
\varphi_{-2}=-\frac{5}{168}(\Delta \alpha)^2.
\ee
The constant coefficients are chosen to keep the same convention of $\varphi_{-2}$ with LVC’s papers \citep{Abbott2019,Collaboration2020,Abbott2016a,Abbott2019c}.
$\Delta \alpha \equiv \alpha_A - \alpha_B$ is the difference between scalar charges of two bodies in a binary. 
For black holes, the no-hair theorem prevents them to acquire scalar charges \citep{Hawking1972,Bekenstein1995,Sotiriou2012,Liu2018}. In many scalar-tensor theories including the models considered in this work, where the no-hair theorem can be applied, scalar charges of black holes are 0.
For neutron stars, scalar charges can be gotten by solving TOV equations.

The detailed process of solving TOV equations to get scalar charges can be found in \citep{Damour1993,Damour1996}. We make a brief review in Appendix \ref{appendix-2} for convenience of reference.
Inputting the explicit form of $A(\varphi)$ and $\alpha(\varphi)$, the EoS and the initial conditions to the TOV equations, one can get the physical quantities $\alpha_A$, $\varphi_0$ and $m_A$ outputted by equations (\ref{physical_quantities}).
The coupling function $A(\varphi)$ and its logarithmic derivative $\alpha(\varphi)$ are specified by a specific theoretical model, which will be discussed in the after subsections.
For the EoS, considering the constraints given by the measurement of PSR J0030+0451 \citep{Miller2019, Riley2019} and the observation evidence that the maximum mass of neutron star can be above $2M_\odot$ \citep{Antoniadis2013,Cromartie2019,Demorest2010,Fonseca2016,Arzoumanian2018}, we select 4 widely used EoS, \texttt{sly}, \texttt{alf2}, \texttt{H4} and \texttt{mpa1}. The tabulated data of EoS are downloaded from the website\footnote{\url{http://xtreme.as.arizona.edu/NeutronStars/data/eos_tables.tar}}.
To solve the differential equations (\ref{TOV_eqs}), the initial conditions, 
\be\label{init_condition}
\mu(0)=0, \ \nu(0)=0, \ \varphi(0)=\varphi_c, \ \psi(0)=0, \ \tilde{p}(0)=p_c, 
\ee
need to be passed into the differential equations solver.
In practices, the initial conditions are taken at the place nearby the center to avoid division by zero. 
The initial values of pressure $p_c$ are taken on a dense grid for interpolation.
The initial condition $\varphi_c$ is determined by shooting method. Different $\varphi_c$ are iteratively tried until a value which can derive the desired $\varphi_0$ is found. 
In order to implement Monte Carlo sampling, we need to get the scalar charge at sufficient speed. It is impracticable to solve the TOV equations every time when a likelihood is evaluated. 
Therefore, we take the values of model parameters and $p_c$ on a dense grid, and solve the TOV equations to get the mass and scalar charge previously. When the Monte Carlo sampler is running, a set of model parameters and $p_c$ sampled by sampler is converted to the mass and scalar charge by linear interpolation.
The interpolation results will be presented in the after subsections.

\subsection{Brans-Dicke Theory}

We firstly consider the Brans-Dicke theory which is usually seen as the prototype of the scalar-tensor theories and has been widely studied. 
The Brans-Dicke theory is characterized by a linear coupling function given by
\be \label{coupling_func_BD}
A(\varphi) = \exp \left(-\alpha_0 \varphi\right),
\ee
which lead to a field-independent coupling strength $\alpha(\varphi)=\alpha_0$.
There is another common convention used in literature \citep{Will2014},
\begin{equation} \label{alpha0_omegaBD}
    \alpha_0^2 = \frac{1}{3+2\omega_{\rm BD}}.
\end{equation}

Given the specific form of coupling function (\ref{coupling_func_BD}), we can use the process discussed in the last subsection to get the scalar charge of a neutron star.
Inputting an initial condition (\ref{init_condition}) and an EoS, we can get the numerical solutions of a neutron star structure by integrating the TOV equations (\ref{TOV_eqs}). And the quantities, $\alpha_A$, $\varphi_0$ and $m_A$ can be extracted from the solutions by (\ref{physical_quantities}). The initial condition $p_c$ and the model parameter $\alpha_0$ are taken on a dense grid for facilitating the interpolation. The last degree of freedom is the asymptotic scalar field $\varphi_0$ which is set to 0 and the initial condition $\varphi_c$ is gotten by the shooting method.
In order to reduce the computational burden, we use an interpolated relation $\alpha_A(\alpha_0, m_A)$ in the Monte Carlo sampling. We present the interpolation result of EoS \texttt{sly} as an example in Figure \ref{fig:BD_sly_m-alphaA}, and results for other EoS can be seen in Appendix \ref{appendix-4}.

Another parameter which called sensitivity, $s_A$, is also commonly seen in literature. The sensitivity and the scalar charge are related by \citep{Palenzuela2014,Sampson2014} 
\be
\alpha_A=\frac{1-2s_A}{\sqrt{3+2\omega_{\rm BD}}}.
\ee
Some works, such as \citep{Zhang2017a}, employ $s_A=0.2$ as a convenient approximation. We illustrate this approximation in Figure \ref{fig:BD_sly_m-alphaA} by the gray dashed horizontal line for comparing with the results gotten by solving TOV equations.

\begin{figure}
    \includegraphics[width=\columnwidth]{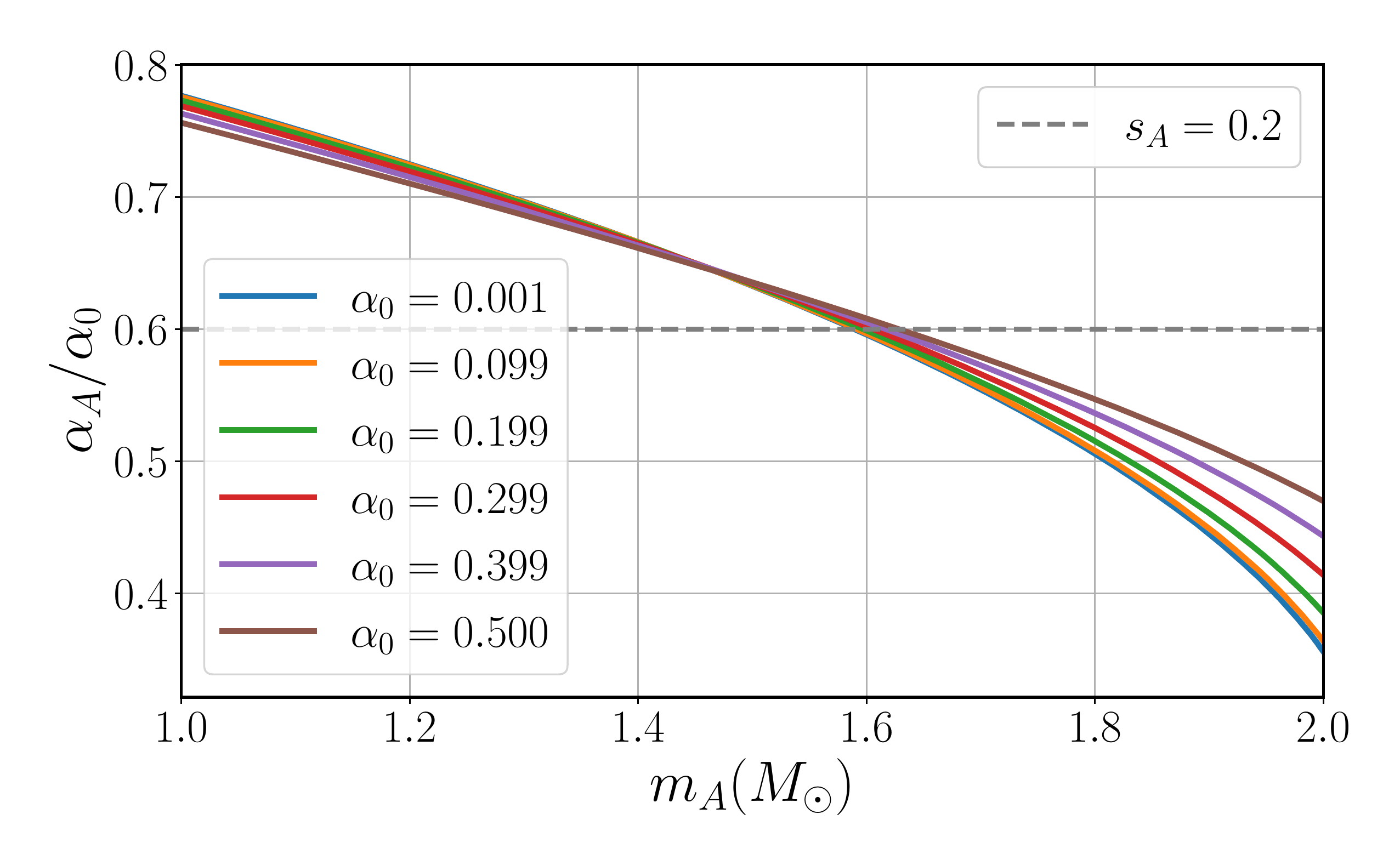}
    \caption{{\bf The interpolation result $\alpha_A(\alpha_0, m_A)$ of EoS \texttt{sly}.} The gray dashed horizontal line denotes the sensitivity $s_A=0.2$ which is a approximation commonly used in literature.}
    \label{fig:BD_sly_m-alphaA}
\end{figure}

\subsection{Theory with Scalarization Phenomena}
In the Brans-Dicke theory, all possible deviations from GR are in the order of $\alpha_0^2$ in both weak-field regimes and strong-field regimes \citep{Will2018,Damour1993}. More generally, in generic scalar-tensor theories, all possible deviations from GR can be expanded as a series of powers of $\alpha_0^2$, which has the schematic form as \citep{EspositoFarese2004,Damour1992}
\be \label{schematic_expansion}
\text{deviation} \sim \alpha_0^2 \times \left[\lambda_0 + \lambda_1\frac{Gm}{R} + \lambda_2\left(\frac{Gm}{R}\right)^2 + ...\right],
\ee
where $m$ and $R$ are mass and radius of a star, $\lambda_0, \lambda_2, ...$ are constant coefficients constructed from $\alpha_0, \beta_0, ...$ in the expansion (\ref{expansion_of_coupling_func}). 
Since the solar system experiments have placed very stringent constraints on $\alpha_0$, it is plausible that all possible deviations from GR in other experiments are expected to be small. The work of Damour and Esposito-Far\`{e}se \citep{Damour1993} shown that such opinions are illegitimate. In the strong-field regime, when the compactness $Gm/R$ excesses a critical value, some nonperturbative effects can emerge, the part of square brackets in expansion (\ref{schematic_expansion}) can compensate the small $\alpha_0^2$, order-of-unit deviations from GR can still be developed.

Following the model discussed by Damour and Esposito-Far\`{e}se in \citep{Damour1993}, we consider the coupling function with a quadratic term,
\be
\ln A(\varphi) = \frac12 \beta_0 \varphi^2.
\ee
The corresponding $\alpha_0$ is given by
\be
\alpha_0 = -\alpha(\varphi_0) = -\beta_0\varphi_0.
\ee
It has been shown in \citep{Damour1993}, when $\beta_0<0$, the local value of $\alpha(\varphi)$ can be amplified with respect to its asymptotic value $\alpha_0$. The nonperturbative amplification effects are expected to take place when $\beta_0\lesssim-4$. 
These nonperturbative amplification effects named spontaneous scalarization can lead to a phase transition in a certain range of mass.
While, if the $\beta_0$ is positive, the deviations from GR are further quenched. In this work, we only consider the negative branch. It returns to GR when $\alpha_0=\beta_0=0$.

\begin{figure}
    \includegraphics[width=\columnwidth]{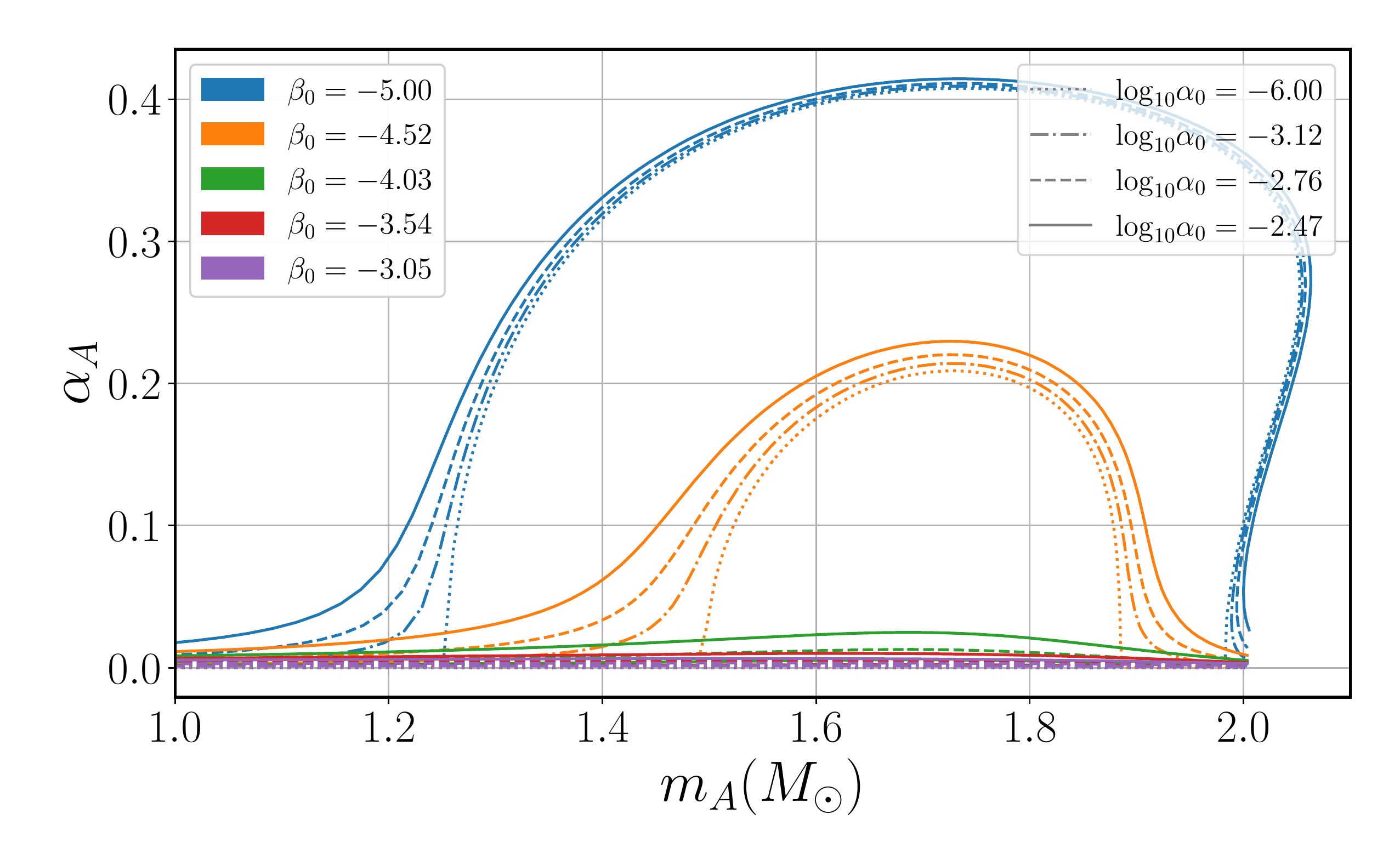}
    \caption{{\bf Nonperturbative strong-field effects in DEF.} The results of one EoS \texttt{sly} are shown as an example. Different colors are used to denote different values of $\beta_0$ and line styles for $\log_{10}\alpha_0$. When $\beta_0<-4$, the nonperturbative strong-field effects emerge, which leads to a phase transition in a certain range of mass and allows the scalar charge to be large even if the $\alpha_0$ is vanishingly small. The nonzero $\alpha_0$ can smooth the phase transition. The varying of the scalar charge as a function of the mass is smoother for larger $\alpha_0$.}
    \label{fig:DEF_sly_m-alphaA}
\end{figure}

The scalar charge of a neutron star can be gotten by solving TOV equations as discussed before. We present the result of the EoS \texttt{sly} in Figure \ref{fig:DEF_sly_m-alphaA} as an example.  
There are two parameters $(\log_{10}\alpha_0, \beta_0)$ characterizing the model in this case. We use colors to denote different values of $\beta_0$, and line styles for $\log_{10}\alpha_0$. The parameter $\alpha_0$ is related to the weak-field effects. We consider the range of $\alpha_0$ under the priori constraint of Cassini $\alpha_0<3.4\times10^{-3}$ \citep{Shao2017,Bertotti2003,Damour2007}. 
Besides, the larger $\alpha_0$ can smooth the phase transition when scalarization phenomena occur \citep{Damour1996}.
For $\beta_0$, it is the parameter that can control whether the spontaneous scalarization could happen in strong-field regimes.
As can be seen in Figure \ref{fig:DEF_sly_m-alphaA}, when $\beta_0<-4$, the scalar charge can be large even if the $\alpha_0$ is vanishingly small for a certain range of mass. 
The mass ranges where the spontaneous scalarization can occur are different for different EoS \citep{Shao2017,Shibata2014}. Therefore, unlike BD, the relations between the scalar charge and the mass of DEF have obvious differences for different EoS. More details will be presented in Appendix \ref{appendix-4}.

\begin{figure}
    \includegraphics[width=\columnwidth]{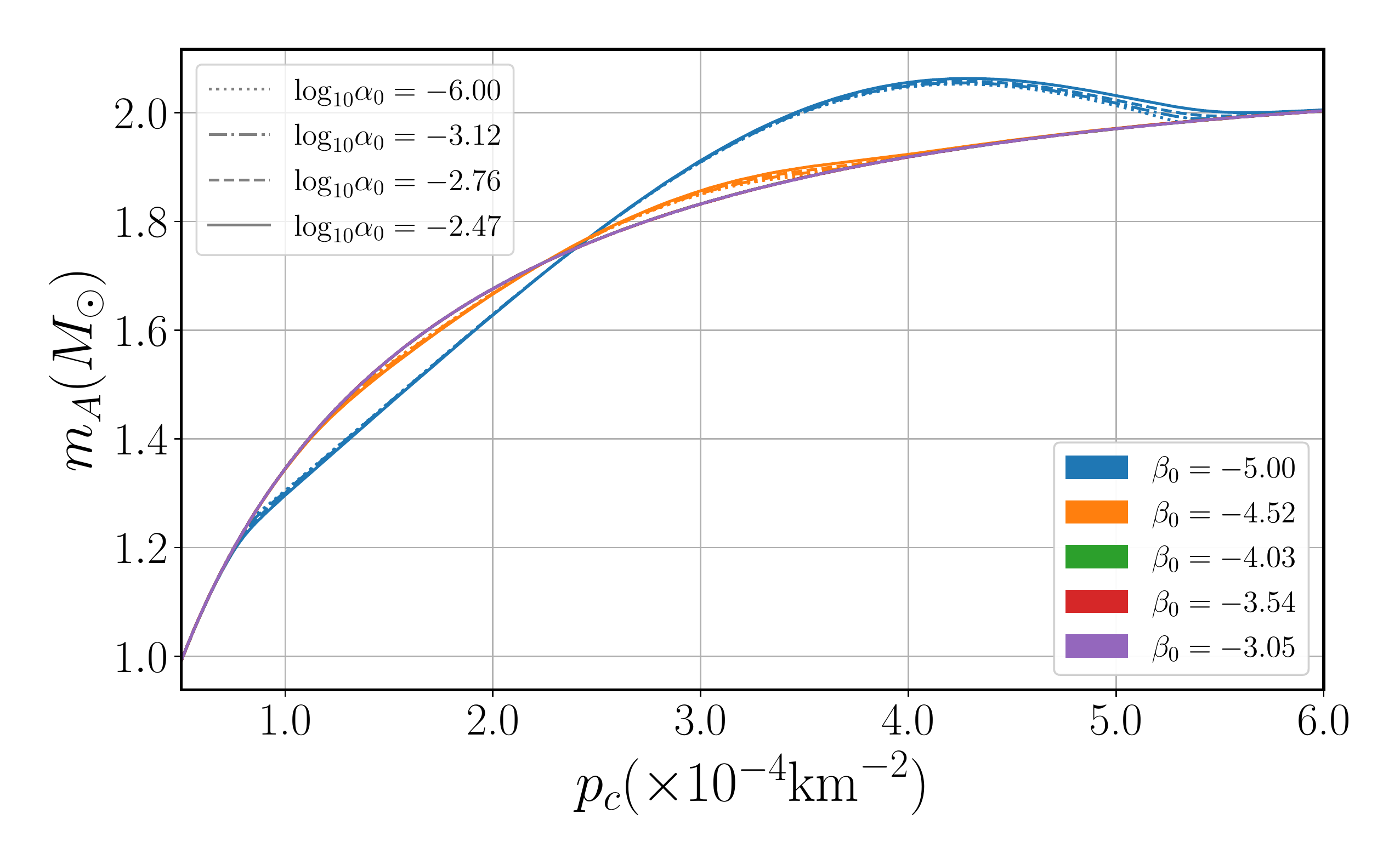}
    \includegraphics[width=\columnwidth]{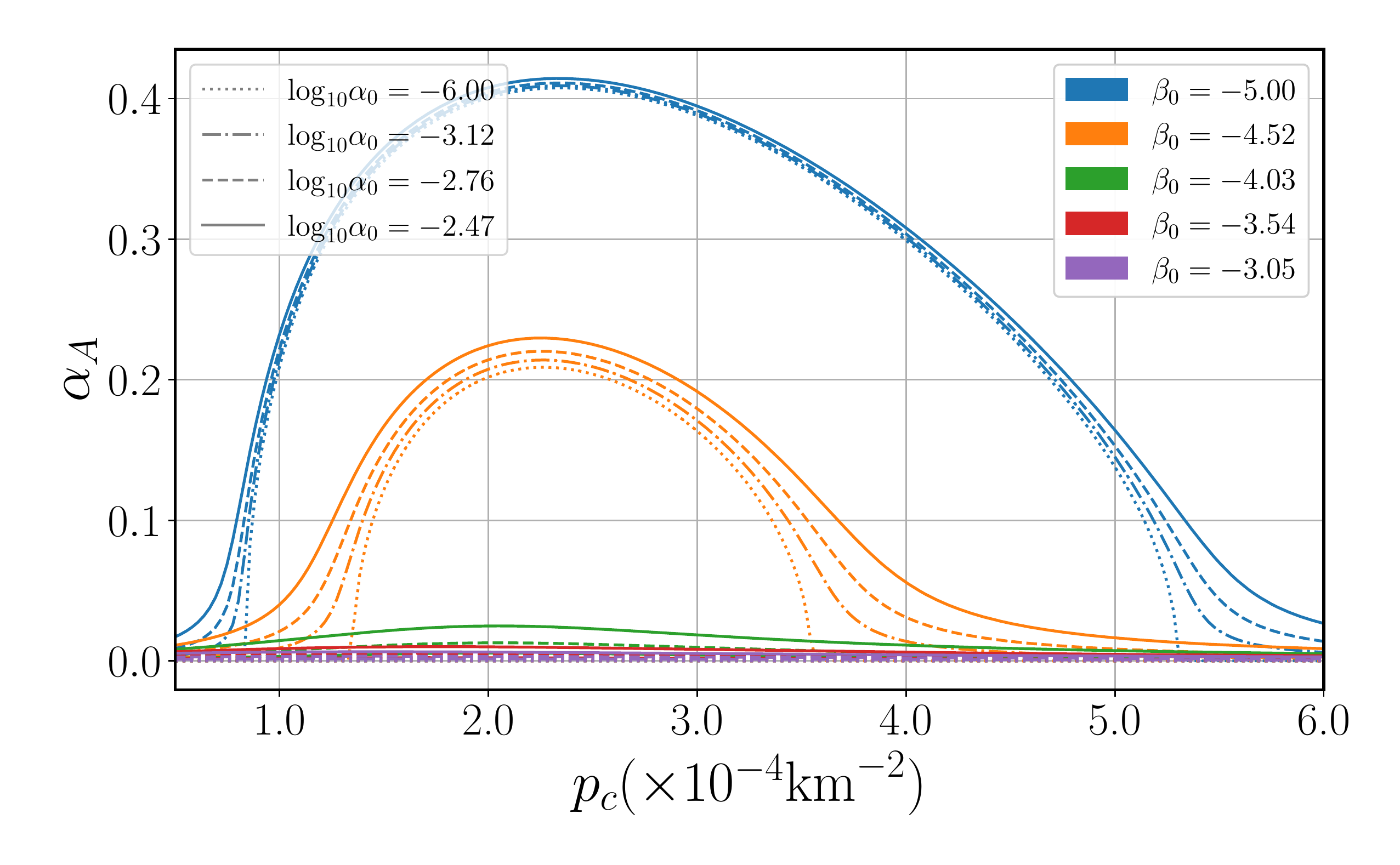}
    \caption{{\bf Interpolation results of $m_A(\log_{10}\alpha_0, \beta_0, p_c)$ and $\alpha_A(\log_{10}\alpha_0, \beta_0, p_c)$ used in stochastic sampling. }We present the results of one EoS \texttt{sly} as an example. Since the relations between the scalar charge and the mass have the hysteresis phenomena, which brings difficulties to interpolation, we use the parameter $p_c$ instead of the mass in the stochastic sampling. The parameters $(\log_{10}\alpha_0, \beta_0, p_c)$ are converted to the parameters $(\alpha_A, m_A)$ which are needed to generate a waveform by the interpolation of $m_A(\log_{10}\alpha_0, \beta_0, p_c)$ and $\alpha_A(\log_{10}\alpha_0, \beta_0, p_c)$ shown in this figure.}
    \label{fig:DEF_interpolation}
\end{figure}

The results gotten by solving TOV equations are interpolated for stochastic sampling. Due to the scalarization phenomena, the curves representing the scalar charge as functions of the mass have the intricate behavior of hysteresis phenomena. In order to facilitating the interpolation, instead of the mass, we use the initial condition of pressure $p_c$ as the parameter sampled in Monte Carlo sampling and generate the interpolation function of $\alpha_A(\log_{10}\alpha_0, \beta_0, p_c)$ and $m_A(\log_{10}\alpha_0, \beta_0, p_c)$. The results of one EoS \texttt{sly} are shown in Figure \ref{fig:DEF_interpolation}.

\subsection{Screened Modified Gravity}
The third model we considered is screened modified gravity (SMG).
Besides the coupling function $A(\varphi)$ characterizing the interaction between the scalar field and the matter field, there is the potential $V(\varphi)$ characterizing the self-interaction of the scalar field. The coupling function $A(\varphi)$ and the potential $V(\varphi)$ define the effective potential $V_{\rm eff}(\varphi)$ which controls the behavior of the scalar field. 
The scalar field acquires the mass around the minimum of the effective potential $V_{\rm eff}(\varphi)$, which depends on the environmental density. The mass of the scalar field can be large in high density regions and the range of the fifth force becomes short, so the effects of the scalar field are screened. While, on large scales, the environmental density is low, the scalar field becomes light and can affect the galactic dynamic or the universe expansion acceleration.
(See comprehensive review \citep{Ishak2018} for more different types of screening mechanism.) For the general SMG with canonical kinetic energy term, we can rewrite the action as 
\begin{equation} \label{action_SMG}
    \begin{aligned}
        S = &  \int{\rm d}^4 x \sqrt{-g_*} 
        \Bigl[\frac{1}{16\pi G}R_* - \frac12g_*^{\mu\nu} \partial_\mu\varphi \partial_\nu\varphi - V(\varphi) \Bigr] \\
        & + S_m\Bigl[\psi_m, A^2(\varphi) g_*^{\mu\nu}\Bigr],
    \end{aligned}
    \end{equation}
where bare potential $V(\varphi)$ characterizes the scalar self-interaction, which endows the scalar field with mass. There are many SMG models in the market, including the chameleon, symmetron, dilaton and $f(R)$ models, in which the functions $V(\varphi)$ and $A(\varphi)$ are chosen as the specific forms \citep{Zhang2016,Liu2018a}. The scalar field equation of motion can be yielded by varying the action with respect to $\varphi$,
\begin{equation}\label{field_eqs_SMG_scalar}
    \Box_{g_*}\varphi  = \frac{\partial}{\partial \varphi} V_{\rm eff}(\varphi), 
\end{equation}
where the effective potential is defined as 
\be
V_{\rm eff}(\varphi) = V(\varphi)- T_*.
\ee

The waveform of gravitational waves from inspiraling compact binaries in SMG has been given in the previous work \citep{Liu2018}. As mentioned above, we only consider the leading order modification which is the dipole term shown in equations (\ref{dipole_term_1}) and (\ref{dipole_term_2}).
Since the effects of the scalar field are suppressed due to the screening mechanism, the scalar charges of neutron stars are expected to be small. Therefore, we do not solve the TOV equations to get the scalar charge, but adopt a simple approximation that considers a neutron star as a static spherical symmetric object with constant density. The scalar field equation (\ref{field_eqs_SMG_scalar}) can be simplified and solved directly to get the exact solution. Matching the internal and external solutions, the scalar charge of a neutron star in SMG can be given by (see Appendix A in \citep{Zhang2017} for more details)
\be \label{scalar_charge_SMG}
\alpha_A = \frac{\varphi_{\rm VEV}}{M_{\rm Pl}\Phi_A}, 
\ee
where $M_{\rm Pl}=\sqrt{1/8\pi G}$ is the reduced Planck mass, $\varphi_{\rm VEV}$ is the vacuum expectation value of the scalar field, and $\Phi_A=Gm/R$ is the surface gravitational potential of the object $A$.

\section{Public Data and Bayesian Method} \label{sec_3}

\subsection{Public Gravitational Wave Data}

{ Among all GW events released by LVC, there are four possible NSBH events \citep{Abbott2021a,Abbott2020,Abbott2020a}, GW190426\_152155, GW190814, GW200105, GW200115, and two (possible) binary neutron star (BNS) events \citep{Abbott2017,Abbott2020b}, GW170817, GW190425, which could potentially be used for the tests on scalar-tensor theories.}

{ For convenience of reference, we list some basic information of these 6 events in Table \ref{data_info_tab} and present a brief review of these events below.}
GW170817 \citep{Abbott2017} is a relatively confident BNS event since its electromagnetic counterpart was captured by various facilities across the electromagnetic spectrum \citep{Abbott2017b}. While definite electromagnetic counterpart observations for the all other events are absent.
For GW190425 \citep{Abbott2020b}, the mass of its components is consistent with neutron stars, but its total mass and chirp mass are larger than those of any other known binary neutron star systems. It cannot be ruled out by GW data alone that one or both of its components are black holes.
GW190814 \citep{Abbott2020a} is a stranger event with its significantly unequal mass ratio and unusual secondary component. It involves a $22.2–24.3 M_{\odot}$ black hole and a $2.50–2.67 M_{\odot}$ object which we do not know much about yet. All current models of formation and mass distribution for compact binaries are challenged by this event.
GW190426\_152155 \citep{Abbott2020} is a possible NSBH event, since the mass of its components is consistent with our current understanding of neutron stars and black holes. But this event has the highest false alarm rate (FAR), $1.4 {\rm yr^{-1}}$, which obscures that whether it is a real signal of astrophysical origin. 
Besides, since the data are uninformative about the effects such as tidal deformability or spin-induced quadrupole, it also cannot be ruled out that its secondary object is a black hole or other exotic objects.
{ GW200105 and GW200115 are two NSBH coalescence events reported recently \citep{Abbott2021a}. The primaries and secondaries of these two binaries have the masses within the range of known black holes and neutron stars respectively. These two events have been regarded as the first observations of NSBH binaries via any observational means. Note that, although the most natural interpretations of these two events are NSBH coalescences, this conclusion is inferred only by their component masses. Until now, there is no direct evidence, such as tidal or spin deformation and electromagnetic counterparts. It is still difficult to rule out that the secondaries are other objects. }

{ Although there are 6 events that probably include at least one neutron star in the current GW catalog, only two events, GW190426\_152155 and GW200115, can be used in this work. }

For GW190814, due to its unusual mass ratio which is in a region that has not been systematically studied, the issue of waveform systematics can lead to some kind of unphysical deviation (referring to Appendix C in \citep{Abbott2020} and Appendix A in \citep{Perkins2021} for more details). And within our knowledge, there are no EoS can reach the mass of its secondary object and meanwhile be favored by current observations of neutron stars. We exclude this event in our discussion.

{ One of two NSBH events reported recently, GW200105, also has a large mass ratio, which can be seen in Figure \ref{mass_ratio_comparision}. The similar unphysical deviation is also present in the analysis of this event. We show the posterior distribution of the dipole modification parameter in Figure \ref{dipole_dev}. The GR value falls in the tail of the posterior and is excluded from the $90\%$ confidence interval. This deviation is believed to be unphysical, which might be the consequences of systematic errors of waveform templates, covariances between parameters, or the way of the parametrization of non-GR modification \citep{Abbott2020,Perkins2021}. 
We present more discussion on this issue in Appendix \ref{appendix-5}, and exclude this event in the main body of this work.
}

For GW170817 and GW190425, only one side limit on the mass ratio can be placed, which means the situation that the two components have an equal mass cannot be ruled out. The dipole radiation depends on the difference of the scalar charges between two components of a binary. As shown in Figures \ref{fig:BD_sly_m-alphaA} and \ref{fig:DEF_sly_m-alphaA}, the scalar charges are functions of mass for BD $\alpha_A(\alpha_0, m)$ and DEF $\alpha_A(\alpha_0, \beta_0, m)$. The symmetrical binaries can lead to very long tails in posterior distributions of $\alpha_0$ or $(\alpha_0, \beta_0)$ which cannot descend to zero when reaching the boundary of whatever prior setting. 
{ Therefore, even the two BNS events can place very strong bounds on the dipole amplitude, we cannot use them to place any effective constraints on model parameters of BD or DEF.}
But we can constrain the dipole radiation for GW170817 and GW190425 without considering specific model parameters.
In order to compare with the results from LVC, we also perform the constraints on $\varphi_{-2}$ for these two events in Appendix \ref{appendix-3}.

Although the origin of GW190426\_152155 still has some uncertainty, the data are consistent with a GW signal from NSBH coalescence. We think it is feasible to test modified gravity models using this event. The results can at least offer a reference for future more confident NSBH events.

As discussed above, the events GW190426\_152155 and GW200115 are the only two left that can be used for our purpose.
The data are downloaded from Gravitational Wave Open Science Center\footnote{\url{https://doi.org/10.7935/99gf-ax93}} \citep{Abbott2021} and down-sampled to 2048Hz.
Besides strain data, power spectral densities (PSDs) are also needed for parameter estimation \citep{Abbott2020c}. Instead of directly estimating PSDs from strain data by the Welch method, we use the event-specific PSDs which are encapsulated in LVC posterior sample releases for specific events \citep{LVC2020,LSCVC2020}. These PSDs are expected to lead to more stable and reliable parameter estimation \citep{Abbott2019,Cornish2015,Littenberg2015}. 
As mentioned above, we only consider inspiral stages, therefore the frequency corresponding to the innermost stable circular orbit (ISCO),
\be \label{f_ISCO}
f_{\rm ISCO} = \frac{1}{6^{3/2}\pi M},
\ee
where $M$ denotes the total mass of the binary, is chosen as the maximum frequency cutoff \citep{Buonanno2009}. 
The minimum frequency cutoffs are chosen by following LVC’s papers \citep{Abbott2020, Abbott2021a}.
The frequency of GW from insprial of compact binary in circular orbit evolves with time.
The data segment durations are set to be consistent with this frequency range.
The data segment is positioned such that there are two seconds post-trigger duration \citep{RomeroShaw2020}.

\begin{table*}
\centering
    \begin{tabular}{llllll}
    \toprule
    {\bf event}                & {\bf type} & {\bf $m_1(M_\odot)$}   & {\bf $m_2(M_\odot)$}   & {\bf SNR} & {\bf FAR$({\rm yr}^{-1})$} \\
    \midrule
    GW170817                   & BNS        & $1.46_{-0.10}^{+0.12}$ & $1.27_{-0.09}^{+0.09}$ & 33.0      & $\le1.0\times10^{-7}$      \\
    GW190425                   & BNS(?)     & $2.0_{-0.3}^{+0.6}$    & $1.4_{-0.3}^{+0.3}$    & 13.0      & $7.5\times10^{-4}$         \\
    ${\rm GW190426\_152155}^*$ & NSBH(?)    & $5.7_{-2.3}^{+3.9}$    & $1.5_{-0.5}^{+0.8}$    & 10.1      & 1.44                       \\
    GW190814                   & NSBH(?)    & $23.2_{-1.0}^{+1.1}$   & $2.6_{-0.09}^{+0.08}$  & 22.2      & $\le1.0\times10^{-5}$      \\
    GW200105                   & NSBH       & $8.9_{-1.5}^{+1.2}$    & $1.9_{-0.2}^{+0.3}$    & 13.9      & 0.36                       \\
    ${\rm GW200115}^*$         & NSBH       & $5.7_{-2.1}^{+1.8}$    & $1.5_{-0.3}^{+0.7}$    & 11.6      & $\le1.0\times10^{-5}$      \\
    \bottomrule
    \end{tabular}
\caption{{\bf Some basic information on 6 events which probably include at least one neutron star are listed for convenience of reference.} The data are copied from Gravitational Wave Open Science Center (\url{www.gw-openscience.org}). The two events with stars are used to place the constraints in this work.}
\label{data_info_tab}    
\end{table*}

\subsection{Bayesian Method}
Bayesian inference is broadly used in modern science for extracting useful information from noisy data. Bayesian inference allows us to make statements on how probabilities of parameters distribute in priori ranges based on the observed data in a specific model. 
In the context of GW astronomy, given a model $M$ described by a set of parameters $\boldsymbol{\theta}$, observed strain data $\boldsymbol{d}$, and background information $I$ which determines the likelihood and prior, the Bayes' theorem can be written as \citep{Abbott2020c,bayes1763lii}
\be
p(\boldsymbol{\theta}|\boldsymbol{d}, M, I) = p(\boldsymbol{\theta}|M, I)\frac{p(\boldsymbol{d}|\boldsymbol{\theta}, M, I)}{p(\boldsymbol{d}|M, I)}.
\ee
The left-hand side is the posterior probability density function of model parameters, which is the product of Bayesian inference and represent the result inferred from data. The three terms on right-hand denote the prior probability density $p(\boldsymbol{\theta}|M, I)$, the likelihood $p(\boldsymbol{d}|\boldsymbol{\theta}, M, I)$, and the evidence $p(\boldsymbol{d}|M, I)$.
Under the assumption that the noise from detectors is stationary and Gaussian, the likelihood function can be written as \citep{Cutler1994, Romano2017} 
\be 
p(\boldsymbol{d}|\boldsymbol{\theta}, M, I) \propto \exp \left[-\frac12 \sum_i \left\langle \boldsymbol{h}(\boldsymbol{\theta})- \boldsymbol{d}| \boldsymbol{h}(\boldsymbol{\theta})-\boldsymbol{d} \right\rangle \right],
\ee
where $i$ denotes different detectors, $\boldsymbol{h}(\boldsymbol{\theta})$ is the waveform template. The angle brackets represent the noise-weighted inner product defined as 
\be
\left\langle \boldsymbol{a}|\boldsymbol{b} \right\rangle = 4 \mathfrak{R} \int \frac{a(f)b^*(f)}{S_n(f)} \ {\rm d}f
\ee
with the noise power spectral density (PSD) $S_n(f)$ of the detector.

As discussed in Section \ref{sec_2}, we consider a waveform model including a term of dipole radiation. 
The waveform template used to compute likelihood is obtained by slightly modifying the aligned-spin with tidal deformability waveform \texttt{IMRPhenomD\_NRTidal} \citep{Dietrich2019} which have been implemented in the LIGO Algorithm Library \texttt{LALSuite} \citep{lalsuite}.

For the prior, a range needs to be set for each parameter of the model. As discussed above, instead of the mass parameter, we choose the central pressure of neutron star as a model parameter. 
The prior ranges are set by referring to \citep{RomeroShaw2020,Abbott2019}. According to the known properties of binary neutron stars, we employ the low-spin prior in this work \citep{Abbott2019a,Stovall2018,Burgay2003}.
The evidence plays the role of the normalization factor and is also used in model selection.

One of the obstacles to applying Bayesian inference is the extremely costly computation. For the huge parameter space, it is impractical to evaluate the likelihood on a grid. The Markov chain Monte Carlo (MCMC) methods \citep{Metropolis1953,Hastings1970} or nested sampling methods \citep{Skilling2006,Skilling2004} are commonly used to estimate the posterior distribution by sampling in parameter space. 
We use the open-source library \texttt{Bilby}\footnote{\url{https://github.com/lscsoft/bilby}} \citep{Ashton2019} with the nested sampler \texttt{Dynesty}\footnote{\url{https://github.com/joshspeagle/dynesty}} \citep{Speagle2020} to do our Bayesian inference. The sampler settings are chosen by referring to \citep{RomeroShaw2020}.

\section{Results and Conclusions} \label{sec_4}
We will present our results in this section. All our results are consistent with GR. For the parameter $\beta_0$ in DEF, we find the constraints given by GWs are comparable with the previous constraints given by pulsar timing experiments. For BD and SMG, the constraints are not competitive with the current bounds placed by the solar system experiments. We do not find significant differences among the constraints using different EoS. More details are in the following.

\subsection{Brans-Dicke Theory}
\begin{figure*}
    \centering
    \includegraphics[width=\columnwidth]{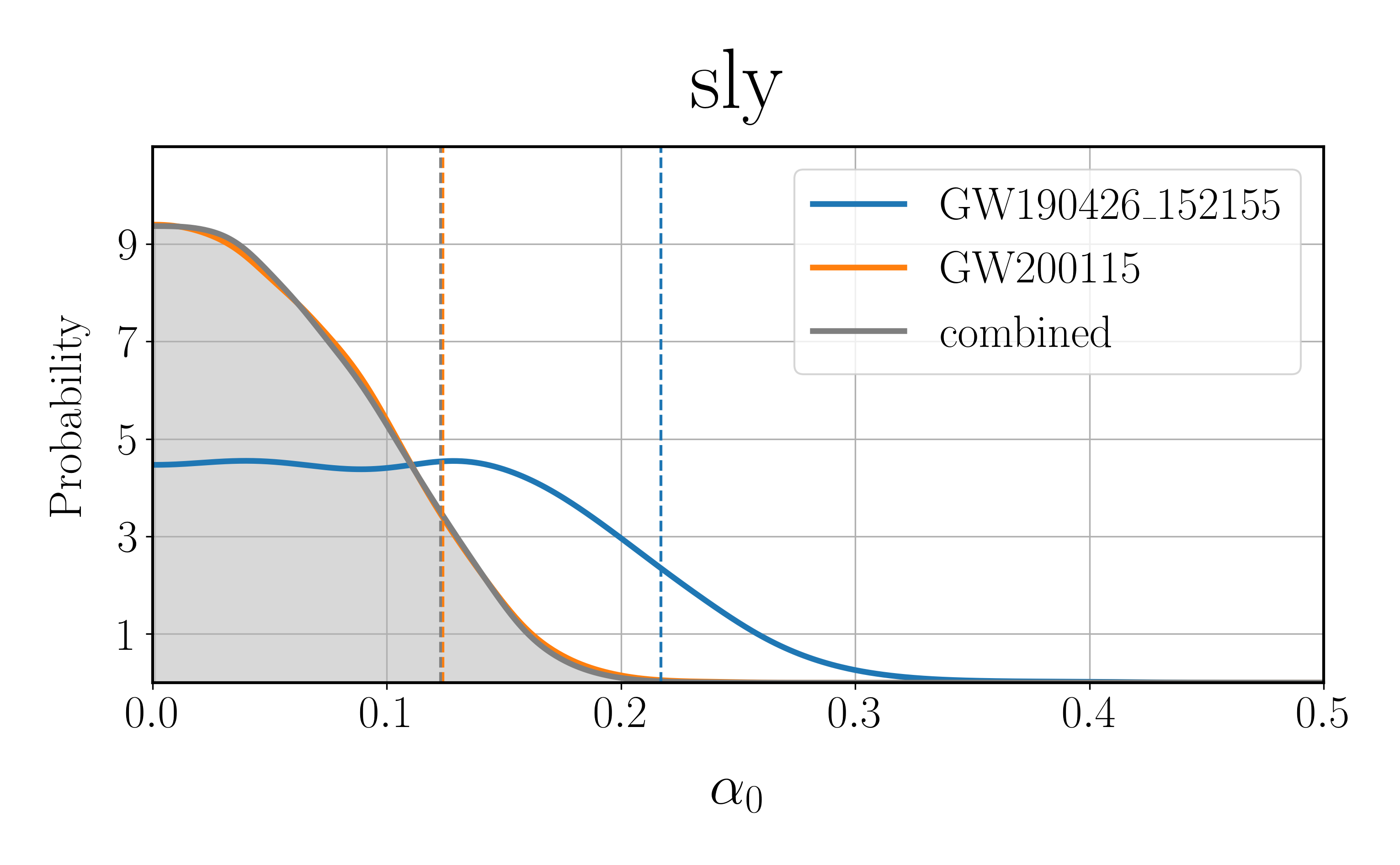}
    \includegraphics[width=\columnwidth]{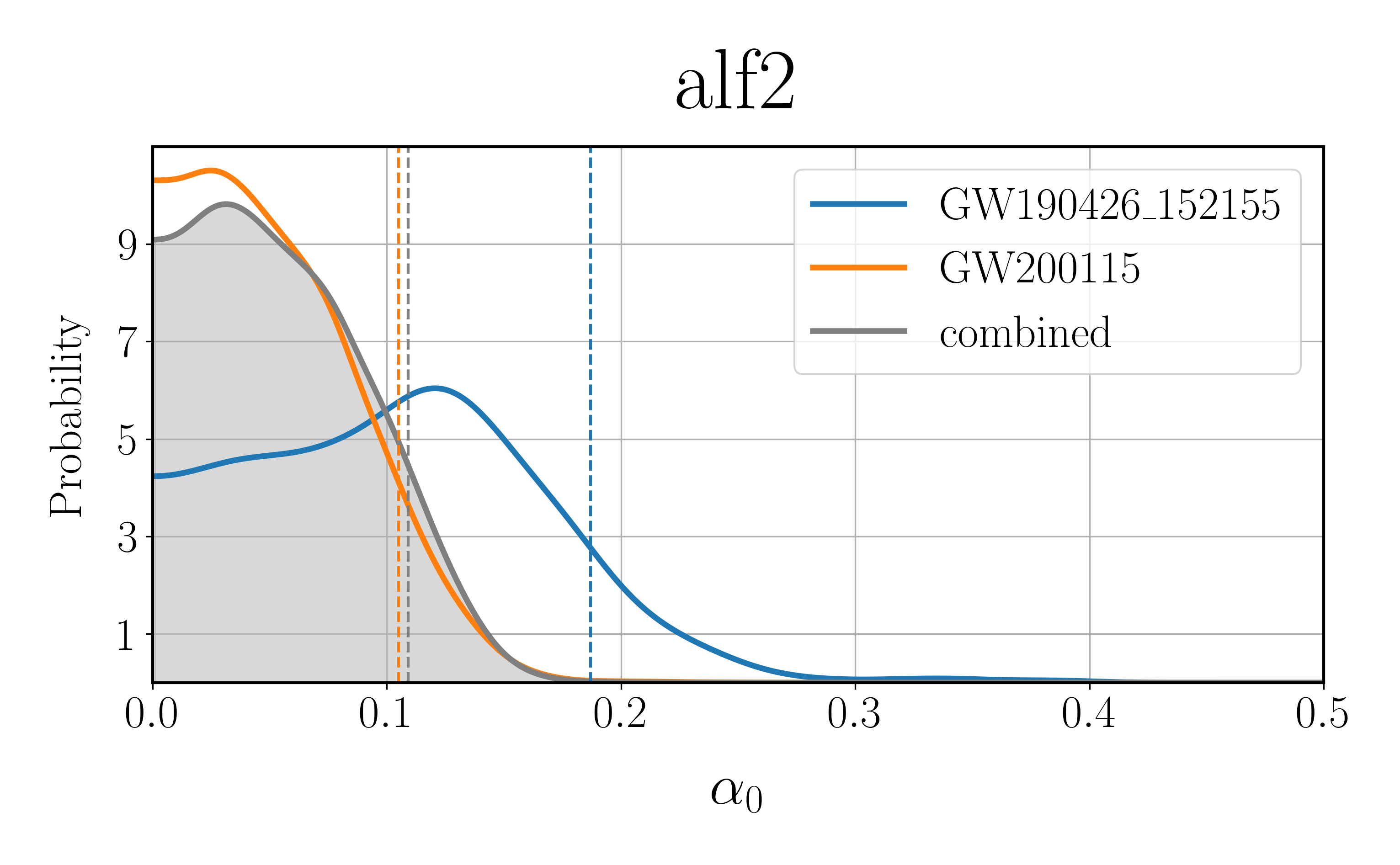}
    \includegraphics[width=\columnwidth]{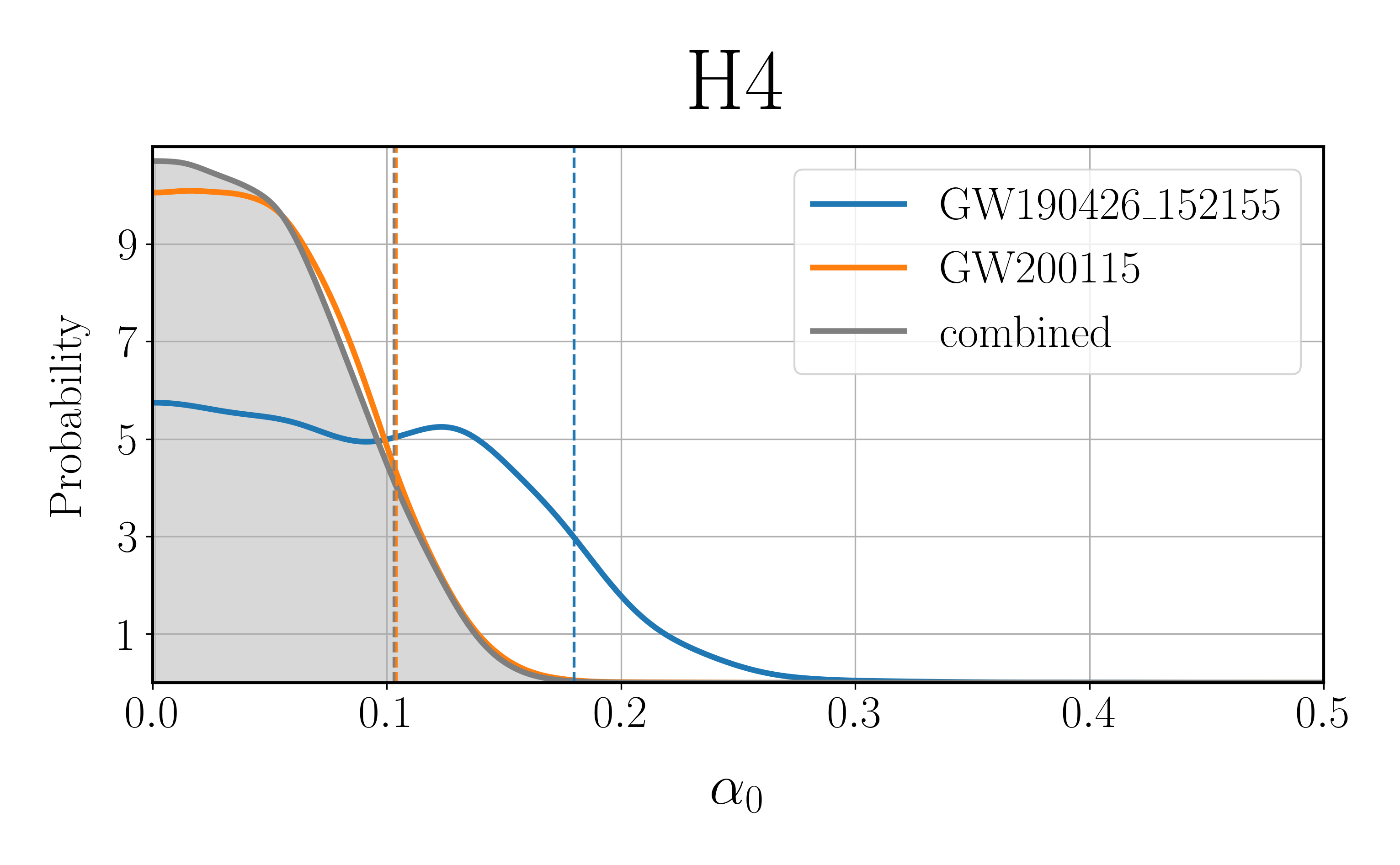}
    \includegraphics[width=\columnwidth]{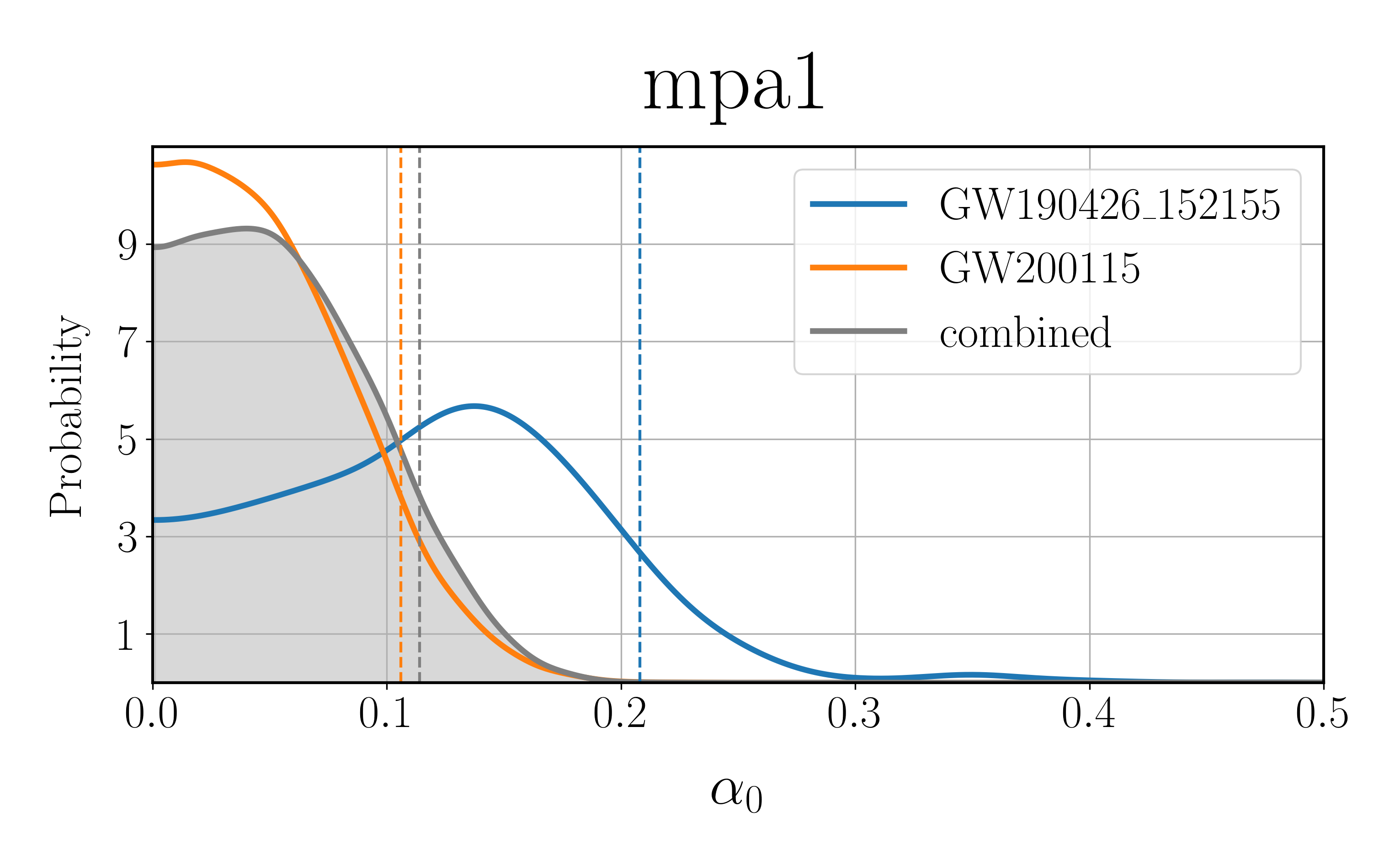}    
    \caption{{\bf Posterior distributions of $\alpha_0$ for BD.} 
    {The results of two events are shown by blue and orange lines. The gray lines with translucent shading denote the combined posterior distributions. The dashed vertical lines indicate the upper limits at $90\%$ CL.}}
    \label{fig:BD_posterior}
\end{figure*}

For BD, the posterior distributions of $\alpha_0$ are shown in Figure \ref{fig:BD_posterior}. 
{ The posteriors of two events can be combined together \citep{Agathos2014,Abbott2019b}, and the combined results are shown by the gray lines with translucent shading.}
The vertical dashed lines denote the upper limits of $\alpha_0$ at $90\%$ confidence level (CL) whose exact values are collected in Table \ref{table:results_vaules}. Colors are used to denote two events. In the results, the impact of difference EoS is invisible within statistical errors.
According to the relation (\ref{alpha0_omegaBD}), one can get the constraints on $\omega_{\rm BD}$ which are also shown in Table \ref{table:results_vaules}. 
So far, the most stringent constraint on BD is from the measurement of Shapiro time delay performed Cassini spacecraft which places the bound \citep{Bertotti2003},
\be
\gamma^{\rm PPN}-1=(2.1\pm2.3)\times10^{-5}.
\ee
The corresponding constraint on $\omega_{\rm BD}$ is \citep{Will2014}
\be
\omega_{\rm BD}> 40000.
\ee
The pulsar timing experiments also place the constraint \citep{Freire2012,Antoniadis2013,Zhang2019}
\be
\omega_{\rm BD}> 13000.
\ee
We summarize the different constraints in the Table \ref{comparision_different_constraints} for comparison. The constraints given by GWs have no competition with these current constraints. This result is expectable. In \cite{Zhang2017a}, we found that in the third-generation GW detector era, the bound by combining a larger number of GW events is expected to be better than that derived in Solar system. 


\begin{table}
    \centering
    \begin{tabular}{lccc}
    \toprule
    {} & \multicolumn{2}{c}{\bf BD} & {\bf DEF} \\
    \cmidrule(lr){2-3}\cmidrule(lr){4-4}
    {} & $\alpha_0$ & $\omega_{\rm BD}$ & $\beta_0$ \\
    \midrule
    {\bf \texttt{sly}}  & $\la0.123$ & $\ga31.5$ & $\ga -3.93$ \\
    {\bf \texttt{alf2}} & $\la0.109$ & $\ga40.6$ & $\ga -4.00$ \\
    {\bf \texttt{H4}}   & $\la0.103$ & $\ga45.6$ & $\ga -3.77$ \\
    {\bf \texttt{mpa1}} & $\la0.114$ & $\ga37.0$ & $\ga -4.08$ \\
    \bottomrule
    \end{tabular}
    \caption{{\bf The vaules of $90\%$ CL limits of combined posteriors for the model parameter in BD, $\alpha_0$, and its corresponding $\omega_{\rm BD}$, as well as the parameter $\beta_0$ in DEF.}}
    \label{table:results_vaules}    
\end{table}

\subsection{Theory with Scalarization Phenomena}
\begin{figure*}
    \centering
    \includegraphics[width=\columnwidth]{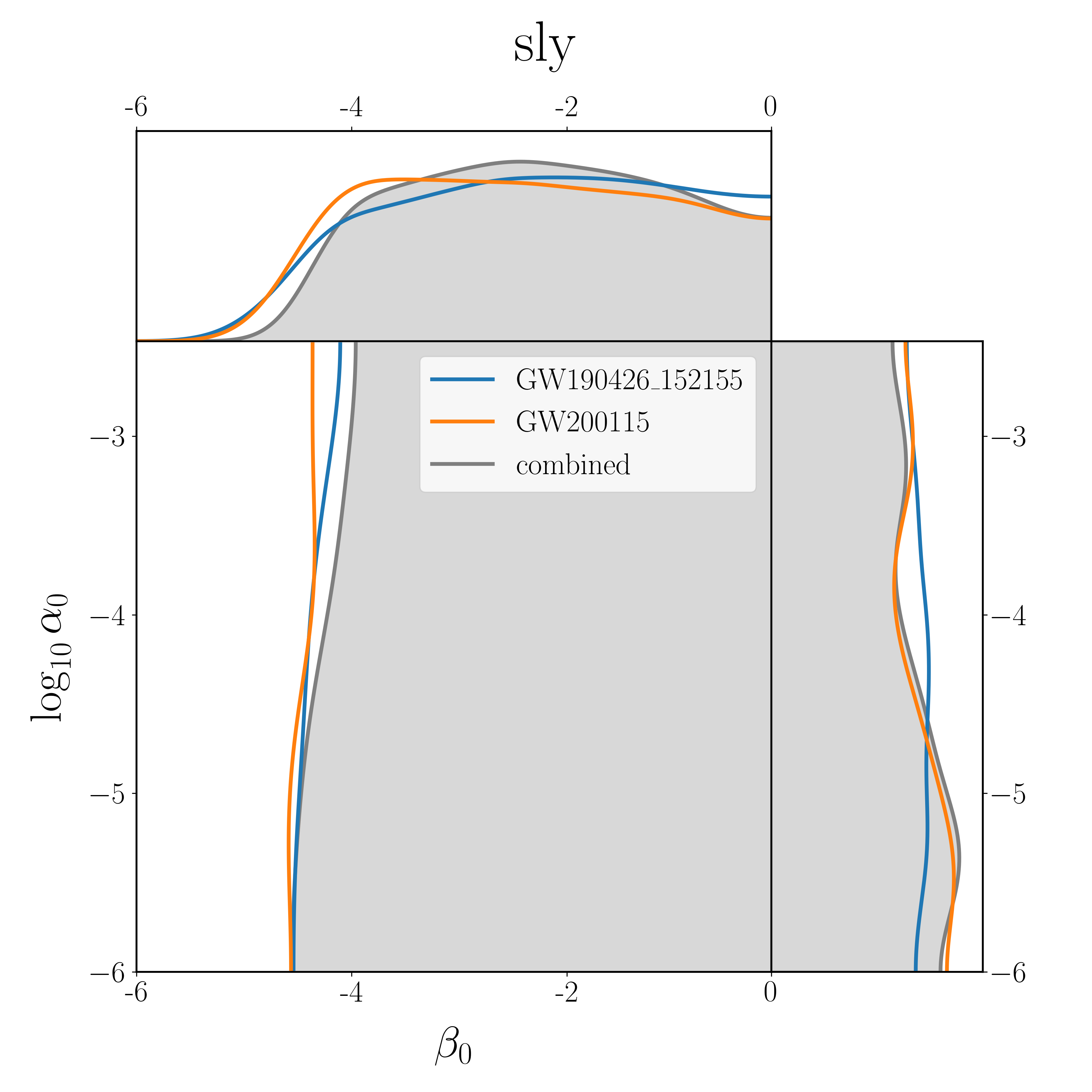}
    \includegraphics[width=\columnwidth]{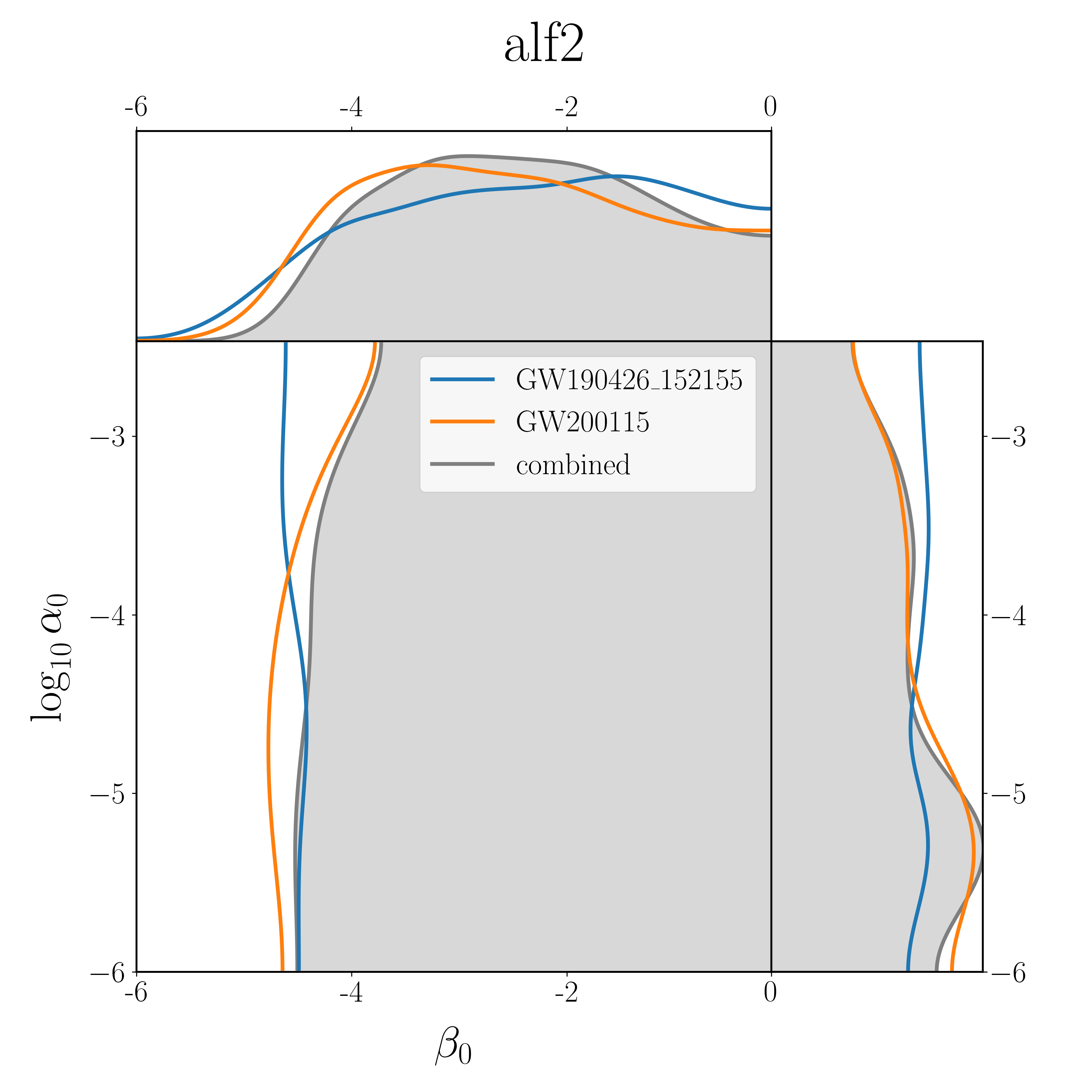}
    \includegraphics[width=\columnwidth]{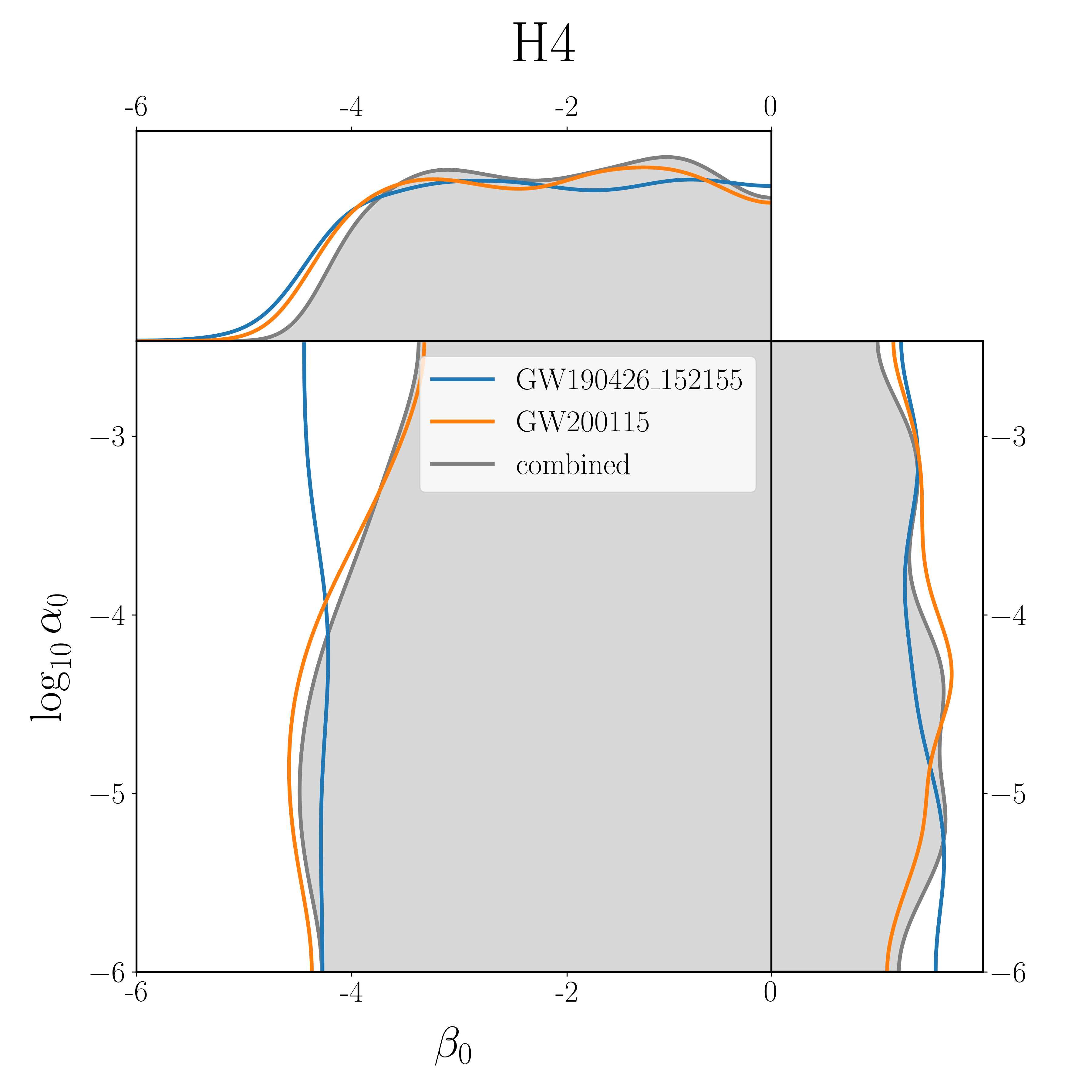}
    \includegraphics[width=\columnwidth]{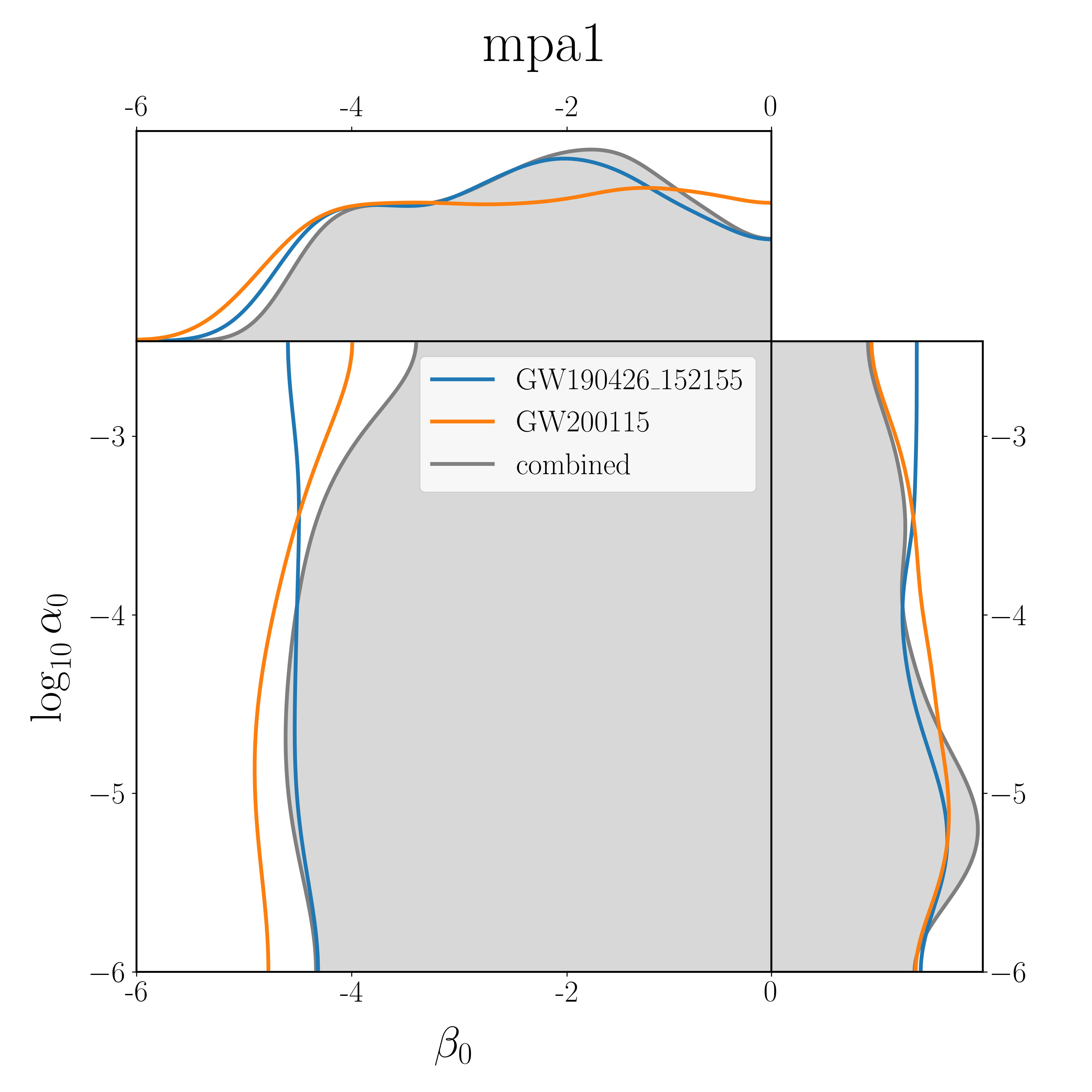}
    \caption{{\bf The posterior distributions of parameters $(\log_{10}\alpha_0, \beta_0)$ in DEF.} 
    { The $90\%$ CL regions of the joint posterior distributions for $(\log_{10}\alpha_0, \beta_0)$ are shown in the main panels, and the marginalized posteriors for $ \log_{10}\alpha_0$ and $\beta_0$ are plotted in the side panels. The posterior distributions of two events are indicated by two different colors and the combined posteriors are shown by the gray lines with translucent shading.}
    It returns to GR when $\alpha_0=\beta_0=0$. The results show consistentcy with GR and no evidence for scalarization phenomena.  
    In the range of $\beta>-4$, the scalarization cannot occur and the scalar charges are almost zero. The different values of $\beta_0$ can hardly be distinguished. Therefore, the distributions of $\beta_0$ are flat in this range.
    Since we require the prior of $\log_{10}\alpha_0$ to be compatible with the Cassini constraint and the influence on the scalar charges of $\log_{10}\alpha_0$ is too small to be detected, the different values of $\log_{10}\alpha_0$ in its prior range are totally indistinguishable by the stochastic sampler.}
    \label{fig:DEF_results}
\end{figure*}

For DEF, we plot the posterior distributions of $(\log_{10}\alpha_0, \beta_0)$ in Figure \ref{fig:DEF_results} and summarize the combined constraints of parameter $\beta_0$ in Table \ref{table:results_vaules}.
In Figure \ref{fig:DEF_results}, we show the $90\%$ CL regions of the joint posterior distributions for $(\log_{10}\alpha_0, \beta_0)$ in the main panels, and the marginalized posteriors for $ \log_{10}\alpha_0$ and $\beta_0$ are plotted in the side panels.
{ The blue and orange lines denote the two events respectively and gray lines with translucent shading denote the combined results. }

Although the mass ranges where the scalarization can occur are different for different EoS \citep{Shao2017,Shibata2014}, we do not find the results have obvious differences beyond statistical errors for different EoS.
It returns to GR when $\alpha_0=\beta_0=0$. Our results are consistent with GR and have no evidence for scalarization phenomena.
There are some features in Figure \ref{fig:DEF_results} which might be noteworthy.

The posterior distributions of $\beta_0$ are almost flat when $\beta_0>-4$. This is because the scalarization phenomena cannot occur in this range.
As can be seen in Figure \ref{fig:DEF_sly_m-alphaA} and the bottom panel of Figure \ref{fig:DEF_interpolation}, in the range of $\beta_0>-4$, the nonperturbative effects will not take place for any neutron star mass. The scalar charges are almost independent of $\beta_0$. Different values of $\beta_0$ can hardly be distinguished by the sampling algorithm. Hence, the posterior distributions are flat in this range.

The posteriors of $\log_{10}\alpha_0$ distributes uniformly on the prior range, which shows no difference with the prior distribution. 
The two-dimensional joint distribution in the main panel of Fiugre \ref{fig:DEF_sly_m-alphaA} also shows that the different values of $\log_{10}\alpha_0$ are totally indistinguishable for the stochastic sampler.
On the one hand, in the range of $\beta_0>-4$ where the nonperturbative amplification effects cannot occur, it returns to the case like the Brans-Dicke theory. And we adopt the prior range compatible with the Cassini constraint, in which the values of $\alpha_0$ are vanishingly small. Any scalar charges evaluated in this region are too small to cause detectable effects. Different values of $\beta_0$ and $\log_{10}\alpha_0$ cannot be distinguished by the sampler in this region.
On the other hand, even in the range of $\beta_0<-4$, as can be seen in Figure \ref{fig:DEF_sly_m-alphaA} and \ref{fig:DEF_interpolation}, the influence on the scalar charges of $\log_{10}\alpha_0$ is much smaller than $\beta_0$. The small difference caused by $\log_{10}\alpha_0$ cannot be detected by the noisy GW data.
Due to these reasons, we cannot place the constraints on the parameter $\log_{10}\alpha_0$ from our sampling results.

It is useful to compare our results with the previous similar works \citep{Shao2017,Zhao2019}, which used pulsar timing experiments to constrain DEF. In the work \citep{Zhao2019}, GW event GW170817 are also considered to place the constraints. Different from the full Bayesian method in which the waveform templates and the power spectral density are used to construct the likelihood function, they employed the measurement results of mass and radii to construct the likelihood.
Another difference is that we use the prior range of $-6<\beta_0<0$ which can return to GR at the edge. While, in the works \citep{Shao2017,Zhao2019}, they are only interested in the range $\beta_0 \in [-5,-4]$ where the scalarization can take place. 
They present the $90\%$ CL bounds
\be
\beta_0 \ga -4.3.
\ee
{ Our constraints of $\beta_0$ are better in the order of $0.3$.}
However, considering the statistical errors, we think this difference is not significant. The different prior setting may also induce this silght difference.
For $\alpha_0$, they can place the constraint $\alpha_0\la10^{-4}$. While the different values of $\log_{10}\alpha_0$ are indistinguishable in our sampling. 
As discussed above, due to the statistical uncertainty and the reason that we consider the prior of $\beta_0$ including the range where the scalarization cannot occur, we cannot constrain $\log_{10}\alpha_0$.
As can be seen in Figure 15 of Appendix A in the work \citep{Zhao2019}, The parameter $\log_{10}\alpha_0$ also cannot be constrained well by using the GW only.
The constraints given by GW170817 in the work \citep{Zhao2019} are a little more related to EoS comparing to our results. 
This is because of the different mass parameters of two GW events. 
The primary and secondary mass of GW170817 with the low-spin prior assumption at $90\%$ CL are given by \citep{Abbott2018,Abbott2019a} $m_1\in(1.36, 1.60)$ and $m_2\in(1.16, 1.36)$. As can be seen in Figure \ref{fig:EoS_DEF}, the scalarization phenomena on these ranges depend on the EoS more strongly. While, the secondary masses of the two events considered here are heavier and in the range where the dependence of scalarization phenomena on EoS is less.
We summarize the comparisons in Table \ref{comparision_different_constraints}

\subsection{Screened Modified Gravity}
The third model we discussed is SMG. As mentioned in Section \ref{sec_2}, the screening mechanism can suppress the effects of the scalar field in high density regions. The scalar charges of neutron stars are expected to be small. Hence, for SMG we do not consider different EoS and strictly solve the TOV equations but adopt a simple approximation which considers neutron stars have a constant density to get the scalar charges as presented in the equation (\ref{scalar_charge_SMG}).
We use the typical value $m=1.4M_\odot$ and $R=10{\rm km}$ for the surface gravitational potential $\Phi_A$ in the equation (\ref{scalar_charge_SMG}).
Since this scalar charge is independent with other parameters under the approximation, we do not sample parameters of specific SMG models. Whereas, we sample the parameter $\varphi_{-2}$ in the equation (\ref{dipole_term_1}) and place the constraint on $\varphi_{\rm VEV}$ by the upper limit of $\varphi_{-2}$.
Constraining the parameter $\varphi_{-2}$ is similar with the model-independent parameterized tests of GW generation performed by LVC \citep{Abbott2019,Collaboration2020,Abbott2019c}, except for two differences.
Since we are discussing the specific model, it is more logical to take physical limits into consideration.
For the models considered in this works, the dipole radiations always take energy away and the outgoing energy flux is positive. The phase evolution will be ahead comparing with the case of GR. So, we consider the prior range constraining $\varphi_{-2}\le0$.
Another difference is that we only consider the inspiral range. Since we are ignorant about the waveform in the merge and ringdown range for scalar-tensor theories, we adopt the cutoff at the frequency corresponding to ISCO as shown in the equation (\ref{f_ISCO}).

\begin{figure}
    \includegraphics[width=\columnwidth]{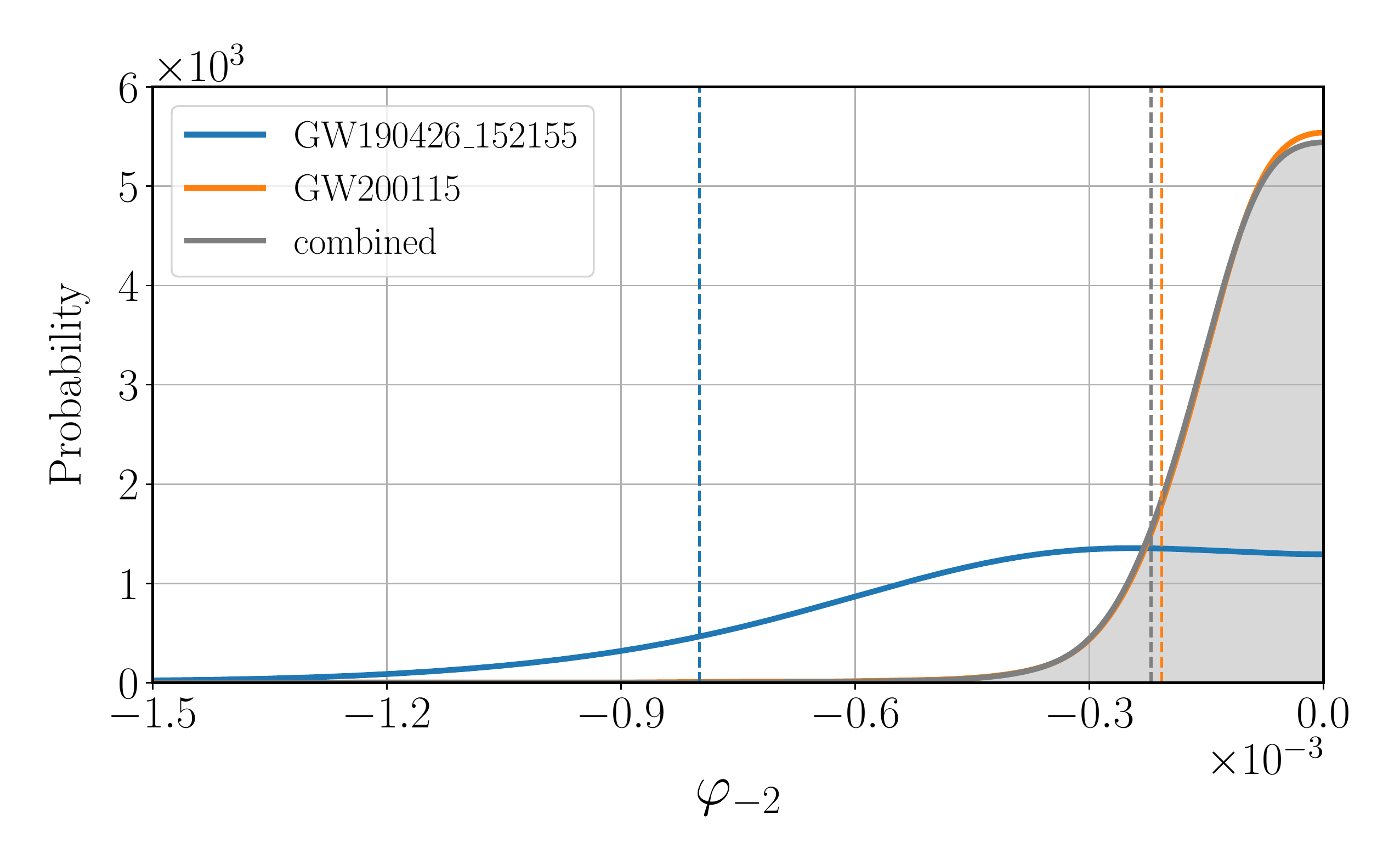}
    \caption{{\bf The posterior distribution of $\varphi_{-2}$.} 
    { As mentioned in the main text, we only consider the physically possible prior range, $\varphi_{-2}\le0$. The blue and orange colors are used to denote the two events, and the combined result is indicated by the gray color. The dashed vertical lines denote the limits at $90\%$ CL.}}
    \label{fig:SMG_result}
\end{figure}

{ The posterior distribution of $\varphi_{-2}$ is shown in Figure \ref{fig:SMG_result}. The combined constraint at $90\%$ CL is }
\be
\varphi_{-2}>-2.2\times10^{-4},
\ee
and the corresponding constraint on $\varphi_{\rm VEV}$ is given by 
\be
\frac{\varphi_{\rm VEV}}{M_{\rm Pl}}<1.8\times10^{-2}.
\ee
The constraint on $\varphi_{-2}$ by GW170817 is about $10^{-5}$ \citep{Abbott2019c} which is one order magnitude better than the constraint given here. This better constraint is because that there are more circles that can be monitored for GW170817. Since GW170817 is lighter than the two events considered here, in the detectors sensitive band the signal can be observed is longer and the circles can be tracked is more. We show the posterior distributions of $\varphi_{-2}$ given by GW170817 and another possible binary neutron star event GW190425 in Appendix \ref{appendix-3} for convenience of comparison.

The parameter $\varphi_{\rm VEV}$ has also be constrained by the solar system tests and pulsar timing experiments. 
The most stringent constraint in the solar system is from lunar laser ranging (LLR) measurement \citep{Hofmann2010,Zhang2019}, which is given by
\be
\frac{\varphi_{\rm VEV}}{M_{\rm Pl}}<7.8\times10^{-15}.
\ee
Pulsar timing experiments also place the constraint \citep{Freire2012,Antoniadis2013,Zhang2019}
\be
\frac{\varphi_{\rm VEV}}{M_{\rm Pl}}<4.4\times10^{-8}.
\ee
These constraints are much better than the constraint gotten in this work. 
On one hand, these much stronger constraints are caused by the fact that the surface gravitational potentials of white dwarfs and objects in the solar system are much less than those in neutron stars. The difference of compact between white dwarfs and neutron stars can be about $\Phi_{\rm WD}/\Phi_{\rm NS} \sim 10^{-4}$. This ratio will be much less for objects in the solar system.
On the other hand, after the GW signal enters the sensitive band, there are only tens of seconds left before the final plunge in. Whereas the pulsar timing experiments can monitor the orbital motion of a binary at lower frequency and in longer time. And the experiments in the solar system can also collect data over long time.

\begin{table*}
    \centering
    \begin{tabular}{lllll}
    \toprule
    {} & {\bf solar system} & {\bf pulsar timing} & {\bf GWs(combined)}  & {\bf GWs(only GW200115)}\\
    \midrule
    {\bf BD} & $\omega_{\rm BD}\ga40000$ & $\omega_{\rm BD}\ga13000$ & $\omega_{\rm BD}\ga40$ & $\omega_{\rm BD}\ga40$ \\
    {\bf DEF} & - & $\beta_0\ga-4.3$ & $\beta_0\ga-4.0$ & $\beta_0\ga-4.2$\\
    {\bf SMG} & $\frac{\varphi_{\rm VEV}}{M_{\rm Pl}}\la7.8\times10^{-15}$ & $\frac{\varphi_{\rm VEV}}{M_{\rm Pl}}\la4.4\times10^{-8}$  & $\frac{\varphi_{\rm VEV}}{M_{\rm Pl}}\la1.8\times10^{-2}$  & $\frac{\varphi_{\rm VEV}}{M_{\rm Pl}}\la1.8\times10^{-2}$ \\
    \bottomrule
    \end{tabular}
    \caption{{\bf Different constraints are summarized for convenience of comparison.} 
    { We also list the results that are given by the events GW200115 only, in case the event GW190426\_152155 is believed to be a false GW signal. However, for BD and SMG, differences between the combined results and the results excluding GW190426\_152155 are within the round-off errors. }}
    \label{comparision_different_constraints}    
\end{table*}

\section{Summary} \label{sec_5}
As more and more various kinds of GW events are observed, GW is becoming an important tool to test GR and explore the nature of gravity. The open access data and user-friendly software tools engage the community to take part in the research about gravitational waves more broadly.
Although various model-independent tests have been performed by LVC and placed stringent upper limits on possible deviations from GR, it is still interesting to ask what constraints on specific models can be placed by last observations.
In this work, we consider three specific scalar-tensor theories, the Brans-Dicke theory (BD), the theory with scalarization phenomena proposed by Damour and Esposito-Far\`{e}se (DEF) and the screened modified gravity (SMG). 

{ The data used in this work are the possible NSBH coalescence GW event GW190426\_152155 in GWTC-2 and one of the two confident NSBH events reported recently GW200115. Due to the possible unphysical deviations, we exclude the events GW190814 and GW200105 in this work. 
Since the dipole amplitude depends on the difference between the scalar charges of two components of a binary, If the possibility that the two components have an equal mass cannot be ruled out, we are unable to place an effective constraint. Therefore, we also exclude the two BNS events GW170817 and GW190425 in the analysis.}

We place the constraints by performing the full Bayesian inference. The waveform template with the dipole term which is the leading order of modification is used to construct the likelihood. The dipole radiation in scalar-tensor theories is proportional to the square of the scalar charge difference between two component objects of a binary. The scalar charges of black holes are zero which is assured by the no-hair theorem. The scalar charges of neutron stars are gotten by solving TOV equations for BD and DEF. For SMG, the effects of scalar field are expected to be small due to the screening mechanism. So, we adopt a simple assumption that the density of a neutron star is a constant to get the scalar charge.

Four tabular EoS are used when solving TOV equations to get the scalar charges for BD and DEF.
However, we do not find the different EoS have significant influences on the results.
All results we get are consistent with GR. 
{ The constraint on BD is about $\alpha_0 \la 0.1$ or equivalent $\omega_{\rm BD} \ga 40$. For DEF, we get the constraint $\beta_0 \ga -4.0$. }
Due to our prior settings and statistical uncertainties, we cannot get the constraints of the parameter $\log_{10}\alpha_0$ in DEF . 
{ For SMG, we place the upper limit $\varphi_{\rm VEV}/M_{\rm Pl} \la 1.8\times10^{-2}$.}
All constraints presented above are at $90\%$.
The constraint on $\beta_0$ in DEF is comparable with the previous constraint from pulsar timing experiments. The constraints on BD and SMG have no competition with previous constraints given by the solar system tests and pulsar timing experiments. 
Although the results of this work do not find any new phenomena or push the current constraints to be more stringent, our results complement the tests on these three specific models in the strong-field regime and make preparations for future more NSBH events.

\acknowledgments
R.N. thanks Yifan Wang for helpful discussions.
This work is supported by NSFC No.11773028, 11633001, 11653002, 11603020, 11903030, 12003029, 11903033, the Fundamental Research Funds for the Central Universities under Grant No.WK2030000036, WK3440000004 and WK2030000044, the Strategic Priority Research Program of the Chinese Academy of Sciences Grant No. XDB23010200, Key Research Program of the Chinese Academy of Sciences, Grant No. XDPB15, and the China Manned Space Program through its Space Application System,
and the China Postdoctoral Science Foundation grant No.2019M662168.
This research has made use of data, software and/or web tools obtained from the Gravitational Wave Open Science Center (https://www.gw-openscience.org/ ), a service of LIGO Laboratory, the LIGO Scientific Collaboration and the Virgo Collaboration. LIGO Laboratory and Advanced LIGO are funded by the United States National Science Foundation (NSF) as well as the Science and Technology Facilities Council (STFC) of the United Kingdom, the Max-Planck-Society (MPS), and the State of Niedersachsen/Germany for support of the construction of Advanced LIGO and construction and operation of the GEO600 detector. Additional support for Advanced LIGO was provided by the Australian Research Council. Virgo is funded, through the European Gravitational Observatory (EGO), by the French Centre National de Recherche Scientifique (CNRS), the Italian Istituto Nazionale di Fisica Nucleare (INFN) and the Dutch Nikhef, with contributions by institutions from Belgium, Germany, Greece, Hungary, Ireland, Japan, Monaco, Poland, Portugal, Spain.

%

\vspace{5mm}
\facilities{LIGO, Virgo}


\software{\texttt{Bilby}\citep{Ashton2019}, \texttt{Dynesty}\citep{Speagle2020}, \texttt{LALSuite}\citep{lalsuite}, \texttt{PESummary}\citep{Hoy2020}, \texttt{NumPy}\citep{Harris2020, Walt2011}, \texttt{SciPy}\citep{Virtanen2020}, \texttt{matplotlib}\citep{Hunter2007}}


\vspace{50mm}

\appendix

\section{Differential Equations for Neutron star Structure} \label{appendix-1}

The scalar charge of a neutron star can be got by solving the TOV equations.
The TOV equations for a neutron star in scalar-tensor theories can be found in previous works \citep{Damour1993,Damour1996}. We present a succinct summary here for the convenience of reference.
Assuming that the neutron star is isolated and nonrotating, the geometry part can be given by the static spherically symmetric metric
\begin{equation}\label{metric}
    \begin{aligned}
    {\rm d}s^2_*  &= g^*_{\mu\nu} {\rm d}x^\mu{\rm d}x^\nu   \\
                  &= -e^{\nu(r)}{\rm d}t^2 + \frac{{\rm d}r^2}{1-2\mu(r)/r} + r^2({\rm d}\theta^2+\sin^2\theta {\rm d}\varphi^2). 
    \end{aligned}
\end{equation}
The matter part is described by the perfect-fluid form of energy-momentum tensor in Jordan frame
\be\label{energy-momentum}
\tilde{T}^{\mu\nu} =  (\tilde{\rho} + \tilde{p})\tilde{u}^{\mu}\tilde{u}^{\nu} + \tilde{p}\tilde{g}^{\mu\nu}.
\ee
We use tilde to denote a quantity in the Jordan frame and star to denote a quantity in the Einstein frame. $\tilde{T}$ and $T_*$ are related by $T_* = A^4(\varphi)\tilde{T}$.
Taking the above matric (\ref{metric}) and energy-momentum tensor (\ref{energy-momentum}) into the field equations (\ref{field_eqs}) and energy-momentum conversation equation $\tilde{\nabla}_\mu \tilde{T}^{\mu\nu} =0$, one can get the following differential equations, which describe the structure of neutron star, 
\be
\begin{aligned} \label{TOV_eqs}
    {\mu}' &= 4\pi G_* r^2 A^4(\varphi) \tilde{\rho} +\frac{1}{2}r(r-2\mu)\psi^2 \\
    {\nu}' &= 8 \pi G_* A^4(\varphi)\frac{r^2}{r-2\mu}\tilde{p} + r\psi^2 + \frac{2\mu}{r(r-2\mu)} \\
    {\varphi}' &= \psi \\
    {\psi}' &= 4\pi G_* A^4(\varphi) \frac{r}{r-2\mu} \left[ \alpha(\varphi)(\tilde{\rho}-3\tilde{p}) + r\psi(\tilde{\rho}-\tilde{p}) \right] \\
    &\mathrel{\phantom{=}} -\frac{2(r-\mu)}{r(r-2\mu)}\psi \\
    {\tilde{p}}' &= -(\tilde{\rho} -\tilde{p})\Bigg[4\pi G_* \frac{r^2A^4(\varphi)\tilde{p}}{r-2\mu} +\frac{1}{2}r\psi^2 + \frac{\mu}{r(r-2\mu)} \\
    &\mathrel{\phantom{= -(\tilde{\rho} -\tilde{p})\Bigg[}} +\alpha(\varphi) \psi \Bigg]. \\    
\end{aligned}
\ee
The above equations can be solved once the EoS, which is the relation between $\tilde{\rho}$ and $\tilde{p}$, and the initial conditions are given.
Physical quantities, the scalar charge, the scalar field at infinity and the gravitational mass, can be extracted from the solution by matching the interior and exterior solutions,
\begin{align}\label{physical_quantities}
    \alpha_A &= -\frac{2\psi_s}{{\nu}'_s} \\
    \varphi_{0} &=\varphi_{s}+\frac{2 \psi_{s}}{\left(\nu_{s}^{\prime 2}+4 \psi_{s}^{2}\right)^{1 / 2}} \mathrm{arctanh}\left[\frac{\left(\nu_{s}^{\prime 2}+4 \psi_{s}^{2}\right)^{1 / 2}}{\nu_{s}^{\prime}+2 / r_{s}}\right] \\
    m_{A} &= \frac{r_{s}^{2} \nu_{s}^{\prime}}{2 G_{*}} \left(1-\frac{2 \mu_{s}}{r_{s}}\right)^{1/2} \exp \left[-\frac{\nu_{s}^{\prime}}{\left(\nu_{s}^{\prime 2}+4 \psi_{s}^{2}\right)^{1/2}} \mathrm{arctanh}\left(\frac{\left(\nu_{s}^{\prime 2}+4 \psi_{s}^{2}\right)^{1/2}}{\nu_{s}^{\prime}+2/r_{s}}\right)\right]
\end{align}
where the subscribe $s$ denotes that the quantities take the values at the star surface.

\section{posterior distribution of other parameters} \label{appendix-2}

\begin{figure}
    \centering
    \includegraphics[width=\columnwidth]{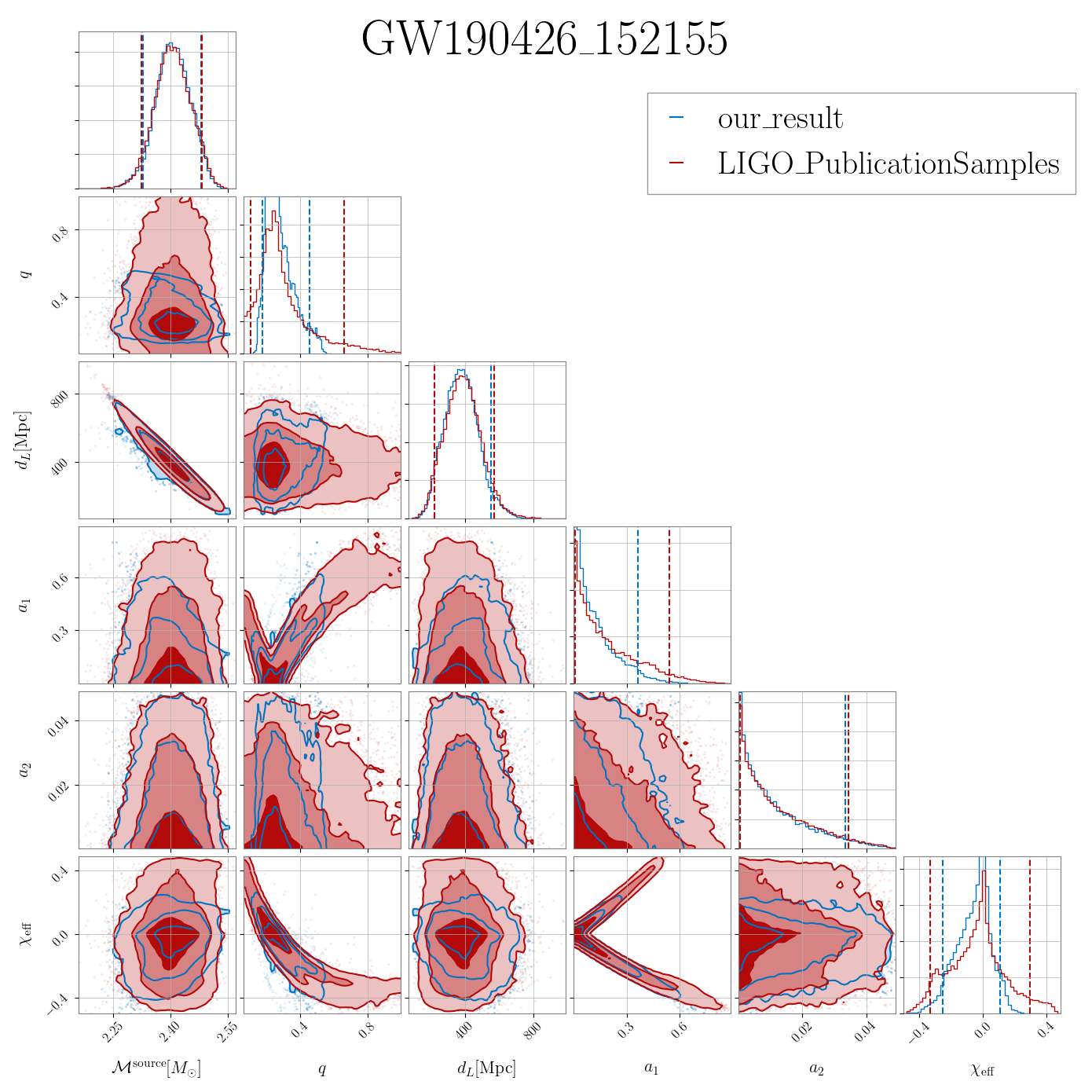}
    \caption{{\bf Comparison between our sampling results and posterior samples released by LVC for the event GW190426\_152155.} The blue regions and lines denote our results and the red for LVC. The dashed vertical lines denote the $5\%$ and $95\%$ quantiles. The labels of parameters follow the conventions in \texttt{bilby}. Since we impose a constraint on the prior of the secondary mass, the distribution of mass ratio and effective inspiral spin parameter have slight differences with LVC. Our sampling results are consistent with the results released by LVC within tolerance.}
    \label{fig:comparision_corner_plot_GW190426}
\end{figure}

\begin{figure}
    \centering
    \includegraphics[width=\columnwidth]{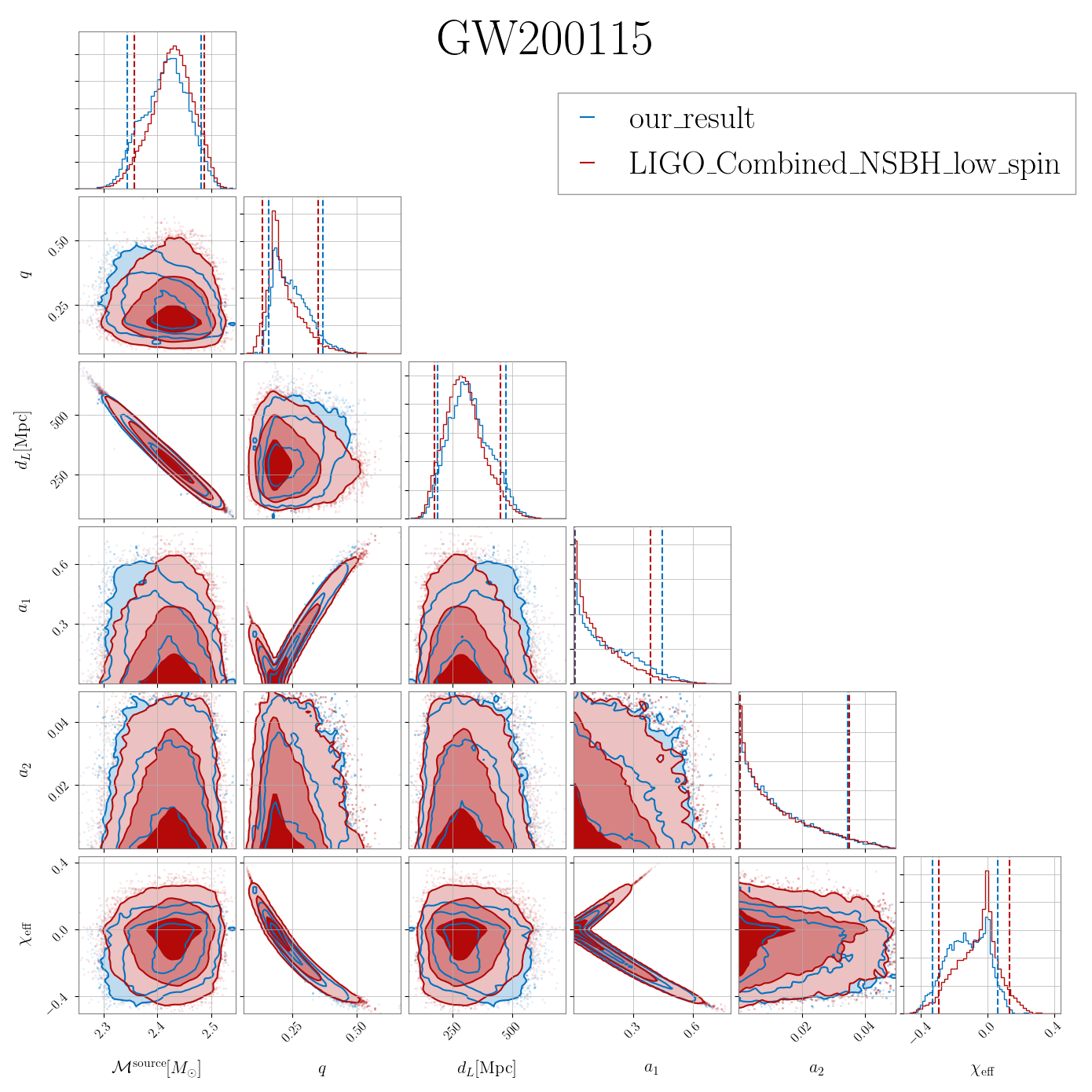}
    \caption{{\bf Comparison between our sampling results and posterior samples released by LVC for the event GW200115.}
    Keeping the same with the last figure, the blue and red colors are used to denote our results and LVC’s, and dashed vertical lines indicate the intervals of $90\%$ CL. The posteriors of GR parameters of the event GW200115 in our runs are also consistent with the results released by LVC within tolerance.}
    \label{fig:comparision_corner_plot_GW200115}
\end{figure}

In order to verify the reliability of our sampling, we compare our results with the posterior data released by LVC\appendixfootnote{\url{https://dcc.ligo.org/LIGO-P2100143/public} for GW200115 and \url{https://dcc.ligo.org/LIGO-P2000223/public} for GW190426\_152155}. We select one of our multiple runs for each event as an example to plot together with parameter estimation samples in the posterior data files released by LVC in Figure \ref{fig:comparision_corner_plot_GW190426} and Figure \ref{fig:comparision_corner_plot_GW200115}. 
The posterior distributions of some interior parameters and the luminosity distance are presented by the corner plot. 

The definitions and labels of the parameters follow the conventions implemented in \texttt{bilby}.
The blue lines and regions denote our results and the red for the results from LVC. The dashed vertical lines represent the $5\%$ and $95\%$ quantiles. 
Since we make our discussion based on the assumption that the secondary of GW190426\_152155 is a neutron star and impose a constraint $m_2 \in [1.0, 2.0]$, the result of mass ratio has slight differences with the result from LVC. Due to the degeneracy between aligned spin and mass ratio, the result of effective inspiral spin parameter has also a little mismatch with LVC result.
Except this, the sampling results of other parameters are consistent with the results released by LVC quite well.

All results of our parameter estimation can be found on Zenodo\appendixfootnote{\url{https://doi.org/10.5281/zenodo.5188445}}.
The differences between all our results and LVC’s are within tolerance.

\section{Comparision of the constraints on dipole radiation} \label{appendix-3}

As discussed in Section \ref{sec_3}, it has practical difficult to constrain $\alpha_0$ or $(\alpha_0, \beta_0)$ by the events GW170817 and GW190425. However, these two events can be used to constrain the dipole radiation without considering specific model parameters.
In order to compare the results given by LVC, we also perform the tests on $\varphi_{-2}$ for these two events. 
We follow the method of model-independent parameterized tests used by LVC \citep{Abbott2019,Collaboration2020,Abbott2016a,Abbott2019c}, except that we only consider the physical range of $\varphi_{-2}<0$ which represents the positive outgoing energy flux.
Following the works of LVC \citep{Abbott2020b,Abbott2019a}, we use the pre-processed data in which the glitches have been subtracted \citep{Driggers2019,Cornish2015,Davis2019,Pankow2018,LVC2017,LVC2018,LVC2019} and event-specific PSDs encapsulated in LVC posterior sample releases \citep{LVC2020a,LVC2019a} to perform full Bayesian inference.
 
\begin{figure}
    \centering
    \includegraphics[width=0.5\columnwidth]{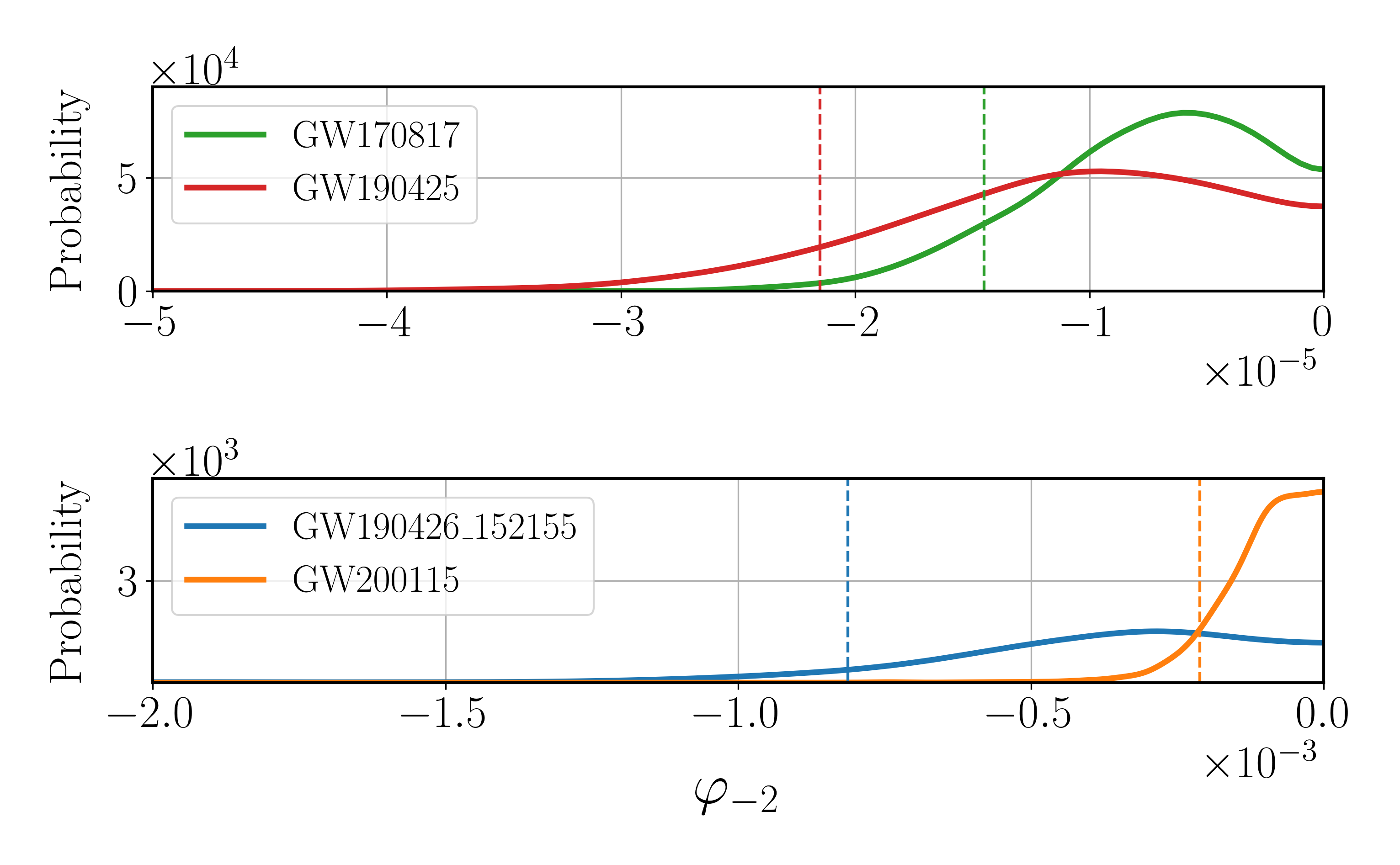}
    \caption{{\bf Comparisons between the posterior distributions of $\varphi_{-2}$.} 
    The dashed vertical lines denote the limits at $90\%$ CL. The limit for GW170817 is about $10^{-5}$ which is consistent with the result reported by LVC \citep{Abbott2019c}. The limit provided by GW190425 is comparable with GW170817, only have a slight difference within the same order of magnitude. While the limits given by two NSBH events considered in this work are much worse than the limits given by the BNS events. The better constraints are because the BNS events have lighter masses which allows more circles of inspiral to be observed in the detectors sensitive band.}
    \label{fig:dipole_plot_comparision}
\end{figure}

The results are shown in Figure \ref{fig:dipole_plot_comparision}.
The limits at $90\%$ CL are shown by the dashed vertical lines. The limit for GW170817 is about $10^{-5}$ which is consistent with the result reported by LVC \citep{Abbott2019c}. The limit provided by GW190425 is comparable with GW170817, only have a slight difference within the same order of magnitude. 
{ While the limits given by the two NSBH events are much worse than the limits given by the two BNS events. The better constraint is because the BNS events has a lighter mass which allows more circles of inspiral to be observed in the detectors sensitive band. Due to the same reason, the limit given by GW200115 is slightly better than limit given by GW190426\_152155.}

\section{relations between the scalar charge and the mass for different EoS} \label{appendix-4}

\begin{figure}
    \centering
    \includegraphics[width=0.5\columnwidth]{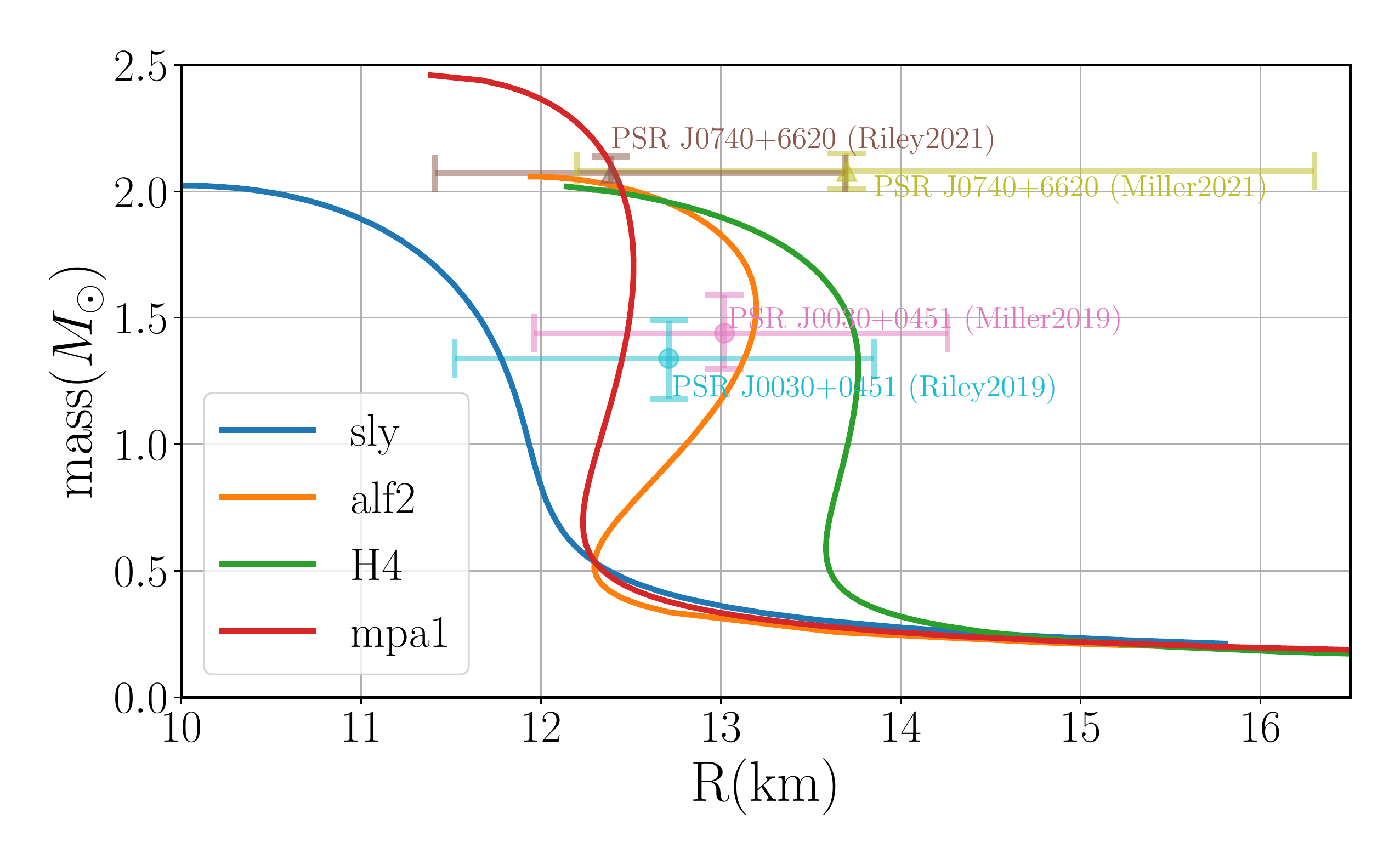}
    \caption{{\bf The relation between mass and radius in GR for four EoS used in this work.} The four solid lines denote the different EoS, and the translucent error bars denote the $68\%$ credible regions of mass-radius measurements of the millisecond pulsar PSR J0030+0451 and PSR J0740+6620. The pink one indicates the result reported in \citep{Miller2019} and the cyan is for the result given by \citep{Riley2019}. The brown and olive are denote the most recent results of PSR J0740+6620 from \citep{Riley2021} and \citep{Miller2021} respectively. We select this four EoS by considering these observation constraints on mass-radius relation and the observation evidence that the mass of a neutron star can excess $2M_{\odot}$ \citep{Antoniadis2013,Cromartie2019,Demorest2010,Fonseca2016,Arzoumanian2018}.}
    \label{fig:EoS_mass-radius}
\end{figure}

The EoS has to be given in order to solve the TOV equation. Considering the measurements of the millisecond pulsar PSR J0030+0451 and PSR J0740+6620 \citep{Miller2019,Riley2019,Miller2021,Riley2021} and observation evidence that the maximum mass of neutron stars can excess $2M_{\odot}$ \citep{Antoniadis2013,Cromartie2019,Demorest2010,Fonseca2016,Arzoumanian2018}, we select four commonly used EoS, \texttt{sly}, \texttt{alf2},, \texttt{H4} and \texttt{mpa1} in this work. We illustrate the relations between mass and radius in GR of these EoS\appendixfootnote{data used to plot are downloaded from \url{http://xtreme.as.arizona.edu/NeutronStars/data/mr_tables.tar}} and the measurements of pulsar mass and radius form two independent groups in Figure \ref{fig:EoS_mass-radius}. The four solid lines represent the EoS used in this work, and the translucent error bars indicate the $68\%$ credible regions of mass-radius measurements.

\begin{figure}
    \centering
    \includegraphics[width=0.45\columnwidth]{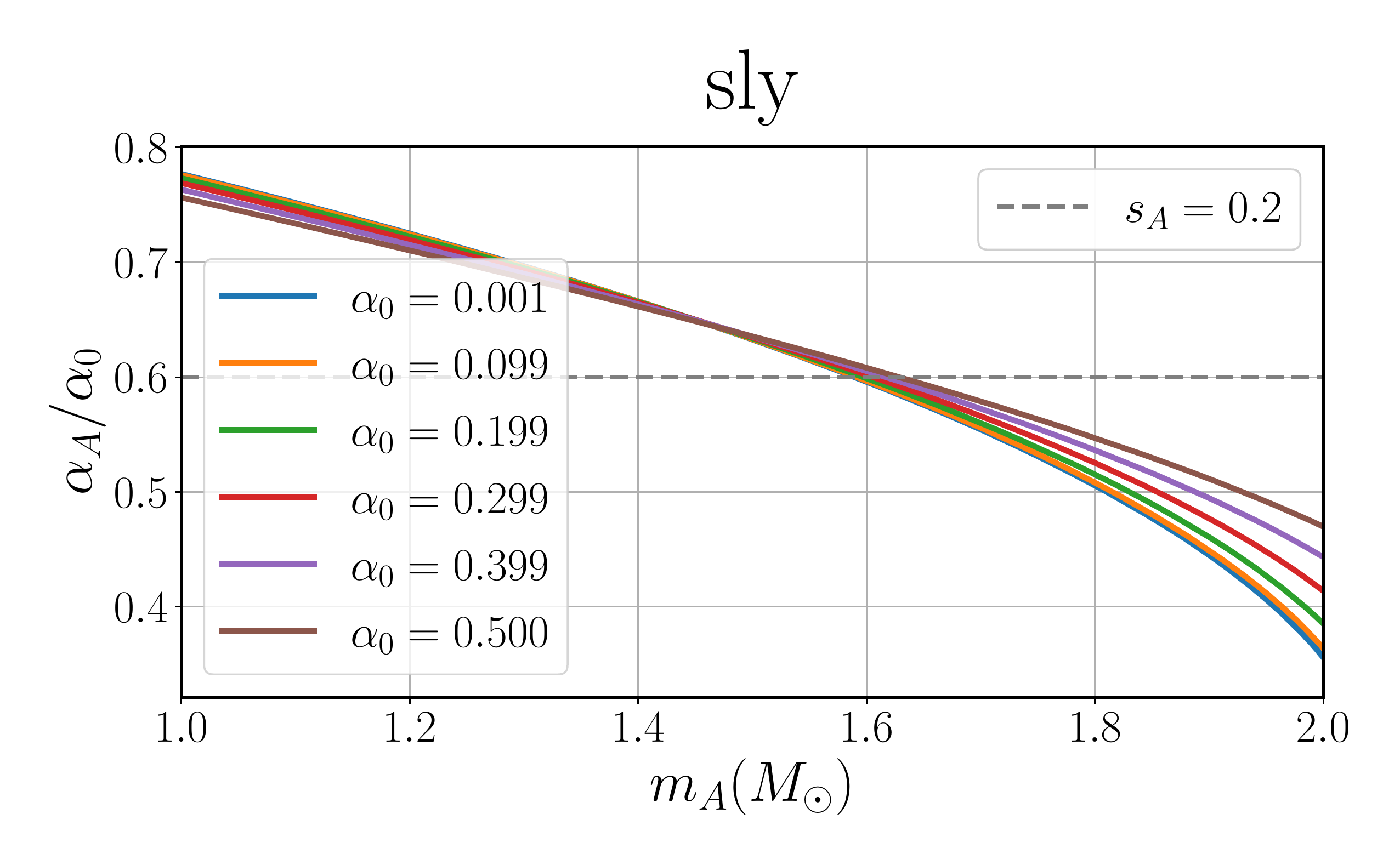}
    \includegraphics[width=0.45\columnwidth]{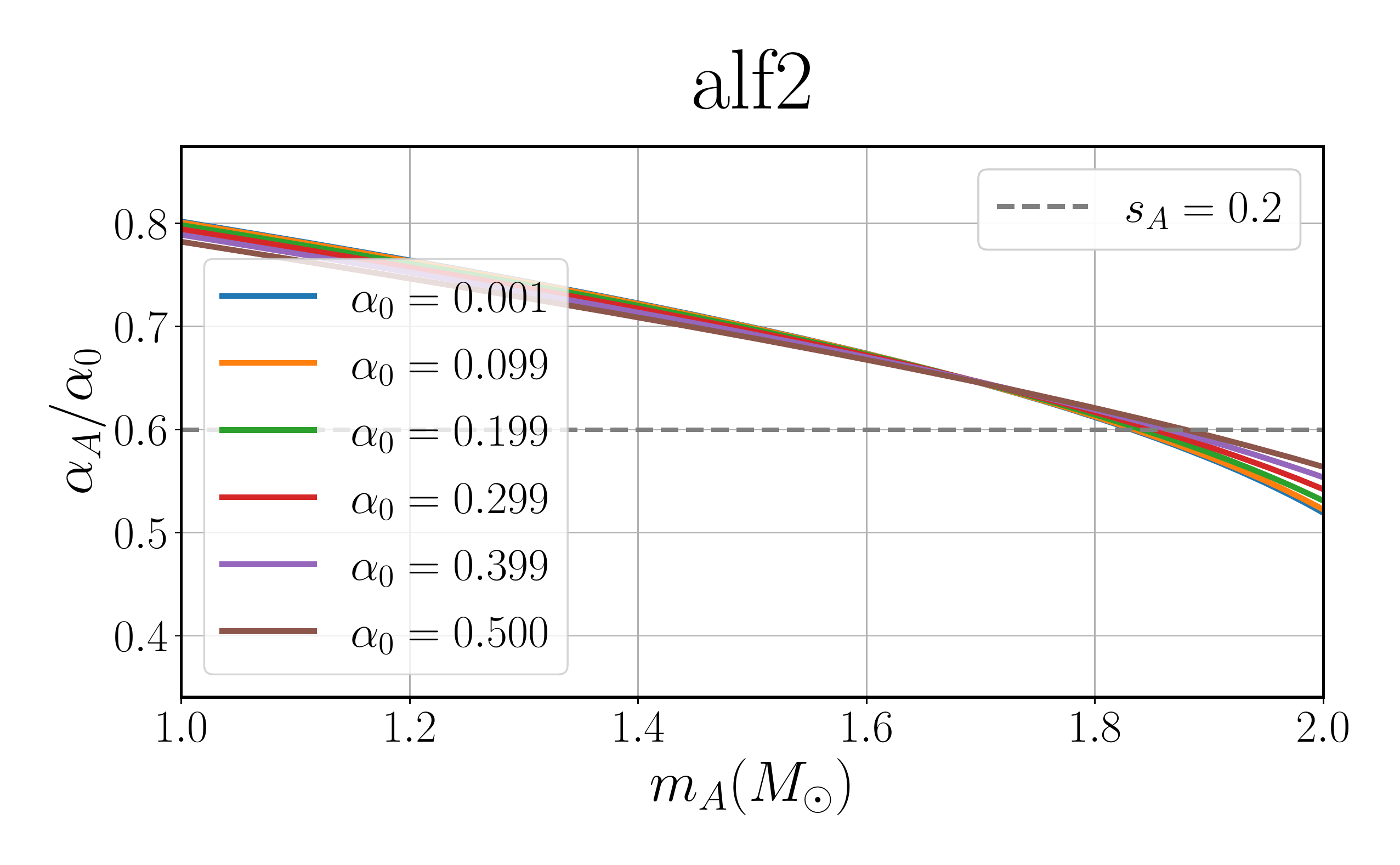}
    \includegraphics[width=0.45\columnwidth]{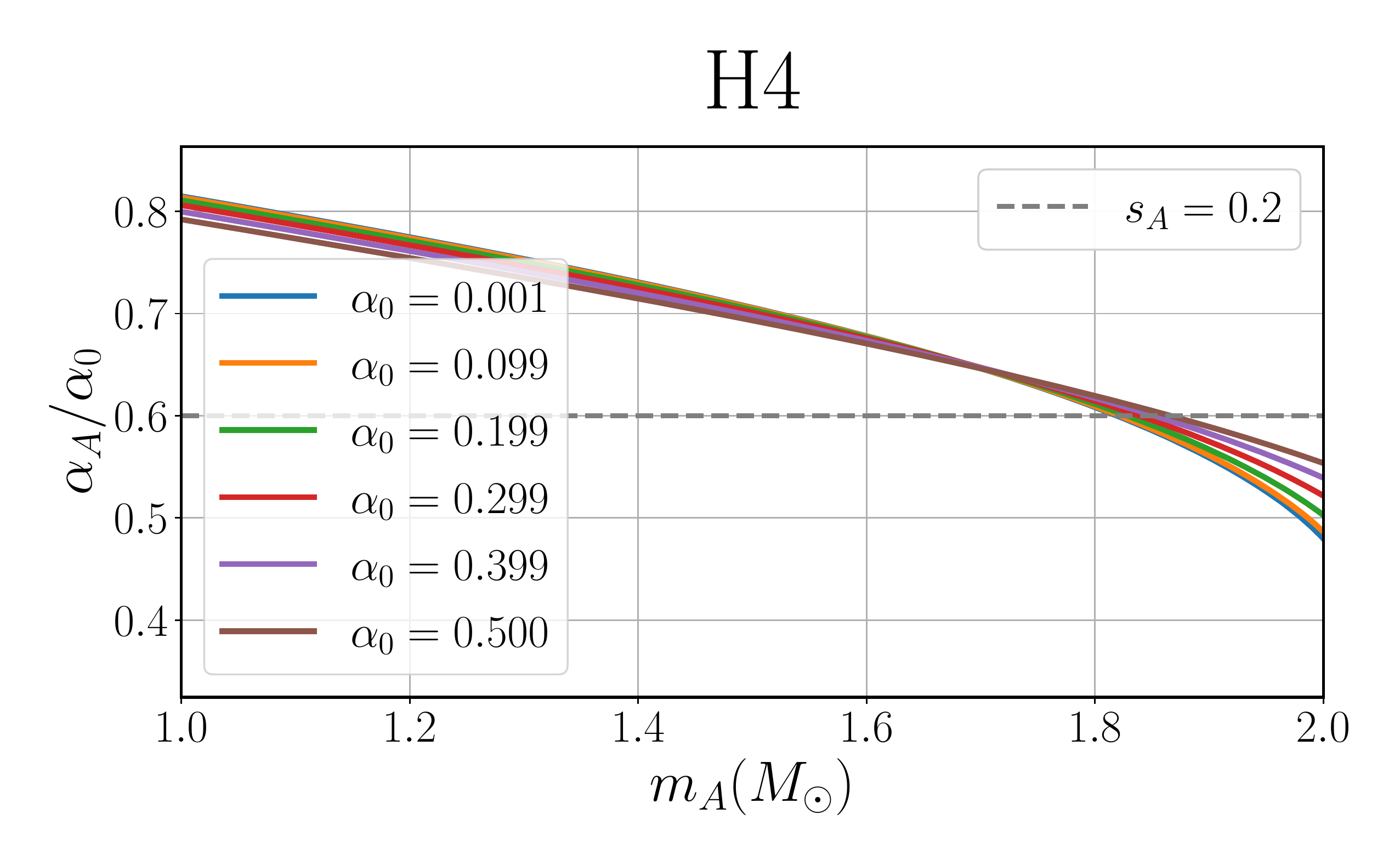}
    \includegraphics[width=0.45\columnwidth]{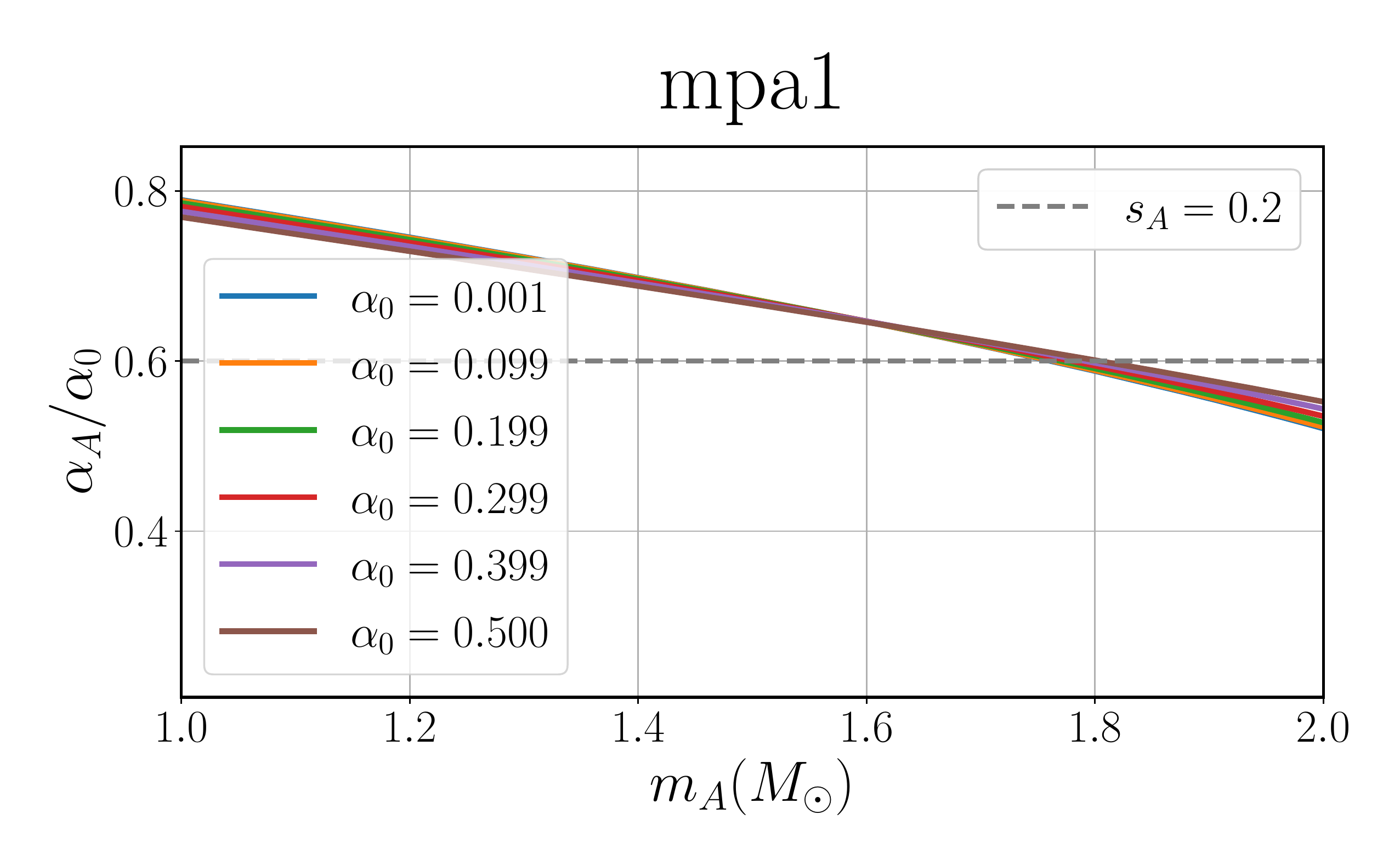}
    \caption{{\bf The relation between scalar charge and mass in BD for four different EoS.} Different colors are used to denote different values of $\alpha_0$. These results show the relation is similar for different EoS in BD. The differences of the scalar charge (relative to $\alpha_0$) are within the order of $0.1$.}
    \label{fig:EoS_BD}
\end{figure}

\begin{figure}
    \centering
    \includegraphics[width=0.45\columnwidth]{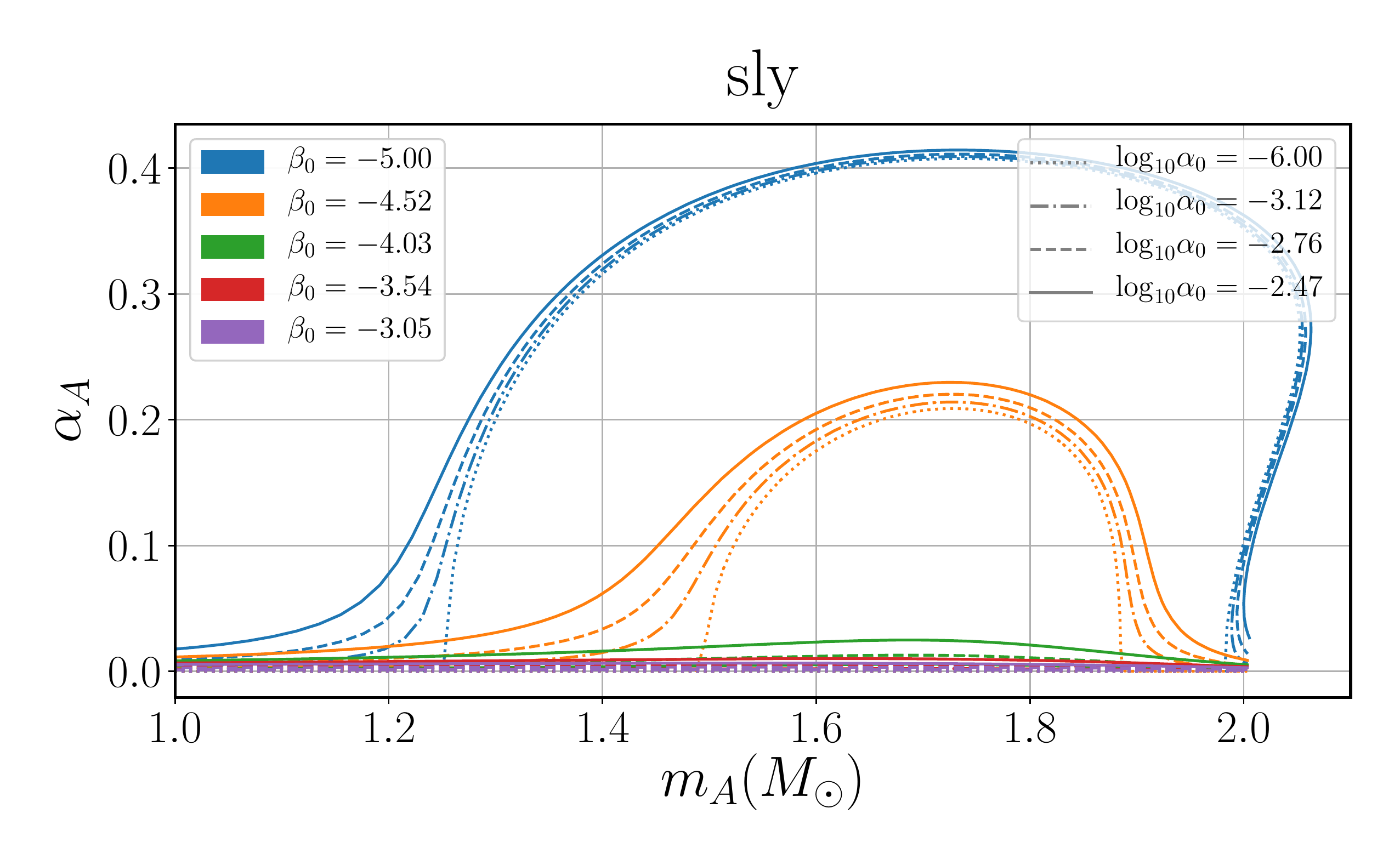}
    \includegraphics[width=0.45\columnwidth]{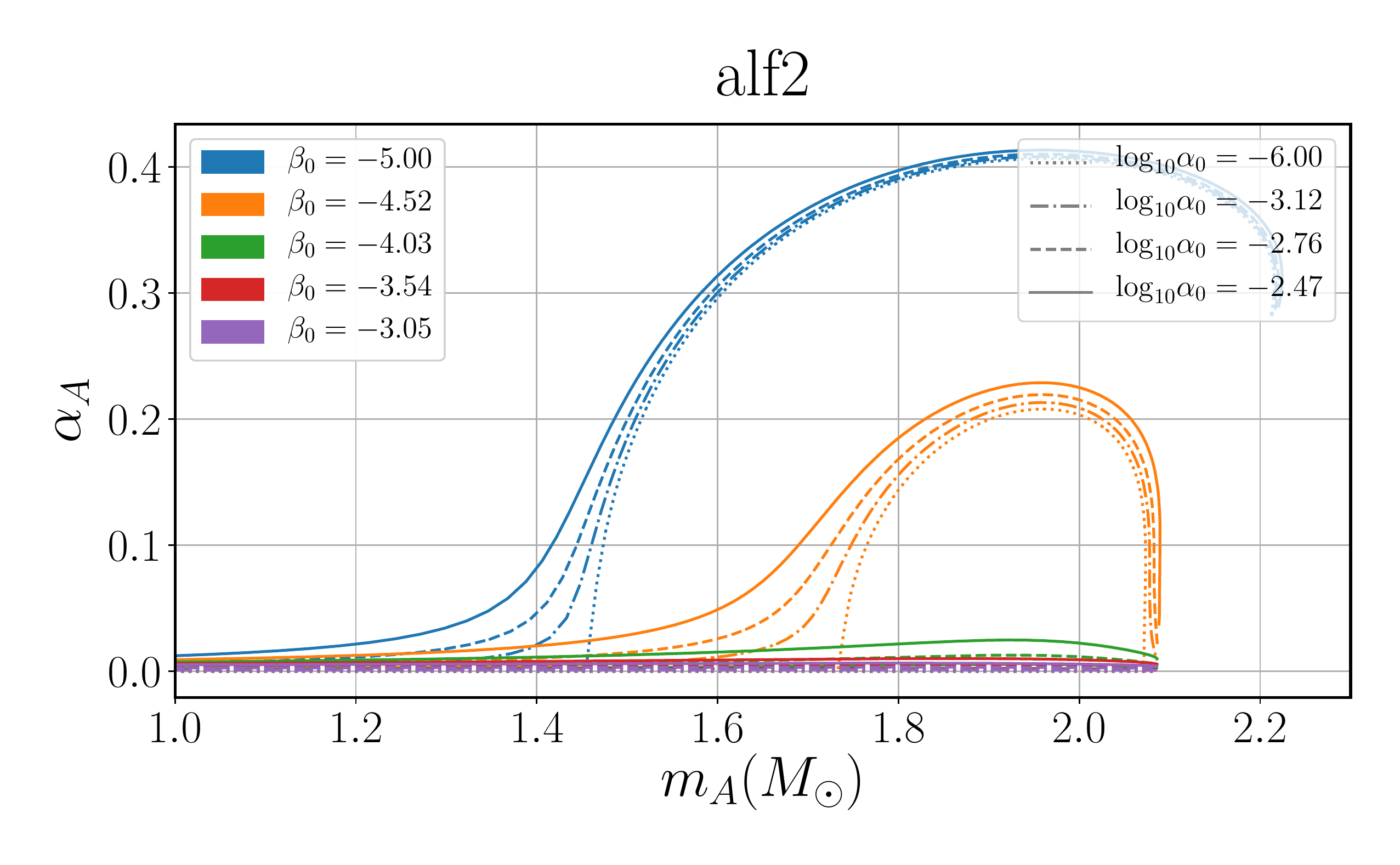}
    \includegraphics[width=0.45\columnwidth]{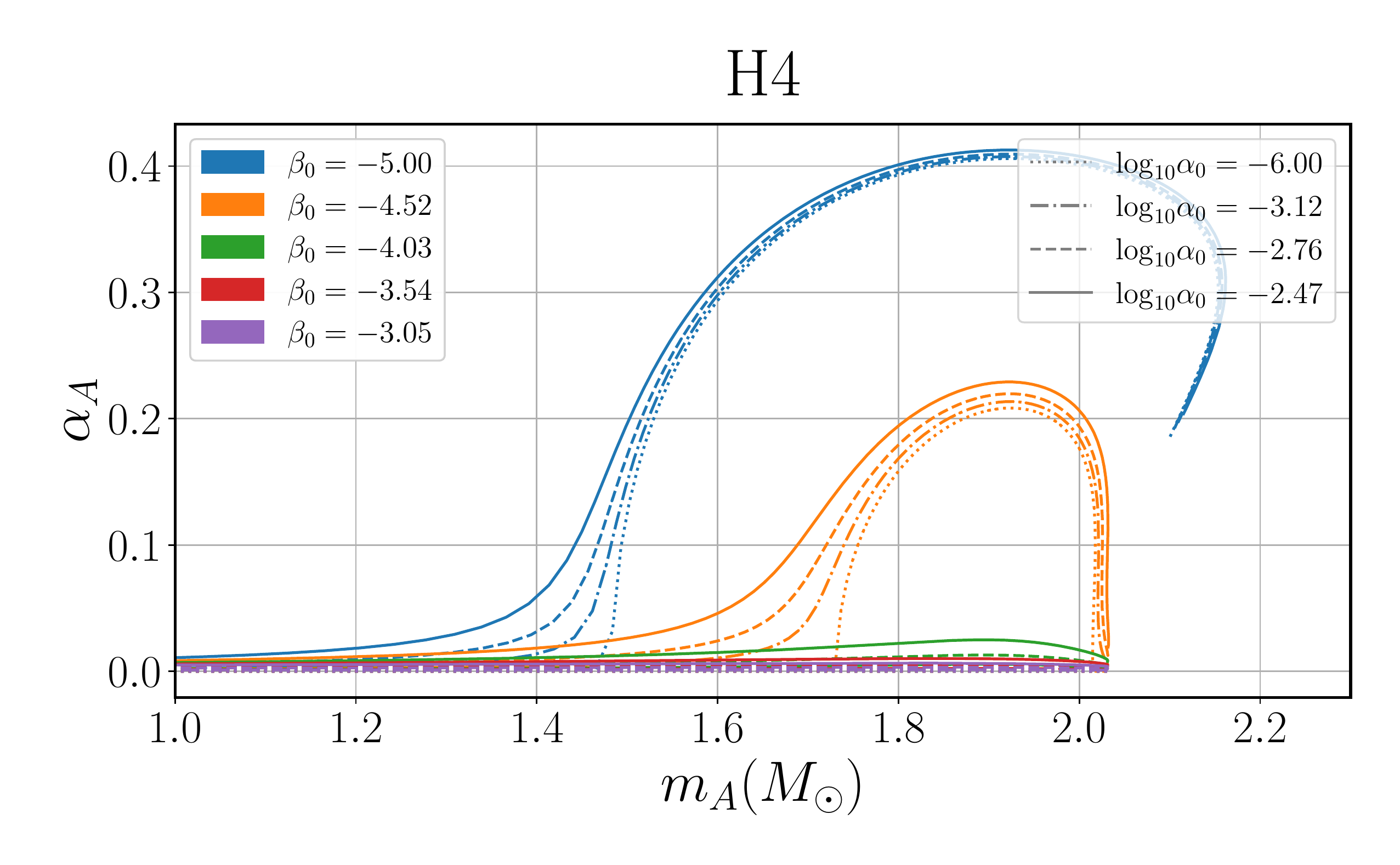}
    \includegraphics[width=0.45\columnwidth]{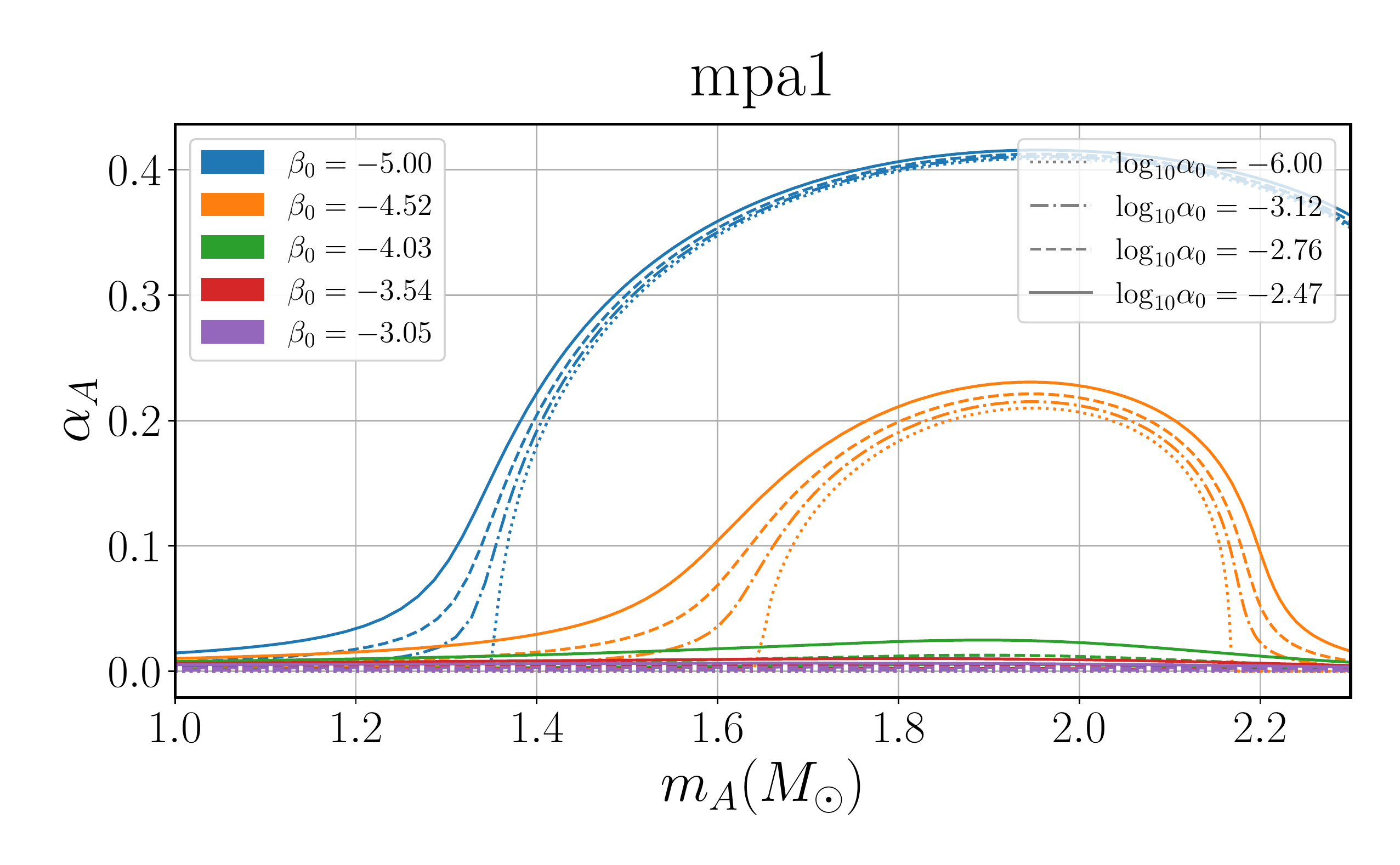}
    \caption{{\bf The relation between scalar charge and mass in DEF for four different EoS.} The colors are used to denote the values of $\beta_0$ and the line styles are used to denote the values of $\alpha_0$.
    For different EoS, the magnitude of scalar charges amplified by scalarization phenomena is almost same, but the scalarization windows which is the mass range where the nonperturbative strong-field effects can occur are different.}
    \label{fig:EoS_DEF}
\end{figure}

Using the four EoS, we can solve the TOV equations by the process discussed in Section \ref{sec_2} and extract the scalar charges and mass from the solutions by the equations (\ref{physical_quantities}). The relations between the mass and the scalar charge are shown in Figure \ref{fig:EoS_BD} for BD and Figure \ref{fig:EoS_DEF} for DEF.
For BD, as can be seen in Figure \ref{fig:EoS_BD}, the influence of using different EoS is slight.
The differences of the scalar charge (relative to $\alpha_0$) are within the order of $0.1$.
Unlike BD, the curves represent the relation between scalar charge and mass have apparent differences for different EoS in DEF. The same conclusion also was presented in previous works \citep{Shao2017,Shibata2014}. For different EoS, the magnitude of scalar charges amplified by scalarization phenomena is almost same, but the scalarization windows which is the mass range where the nonperturbative strong-field effects can occur are different.

\section{Unphysical deviations of highly asymmetric sources} \label{appendix-5}
\begin{figure}
    \centering
    \includegraphics[width=0.5\columnwidth]{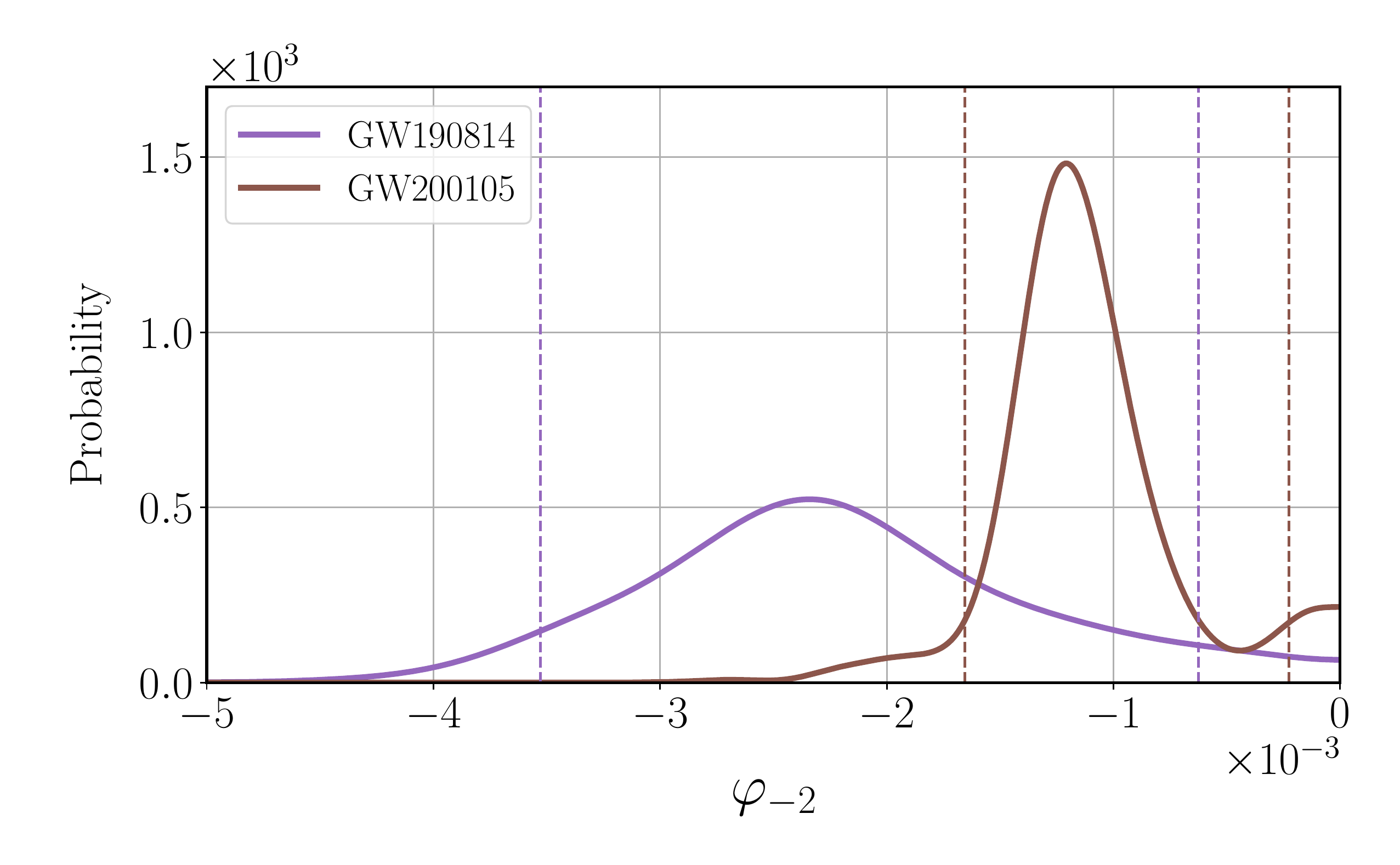}
    \caption{{\bf The posteriors of the dipole modification parameter $\varphi_{-2}$ for the two NSBH events excluded in this work.} 
    The dashed vertical lines indicate $5\%$ and $95\%$ percentiles for the two events respectively. In the posterior of GW190814, the best fit value deviates from the GR value in the order of $10^{-3}$, the GR value falls in the tail and is excluded from the $90\%$ confidence interval. The deviation shown here is in agreement with the LVC analysis which can be seen in Figure 19 of \citep{Collaboration2020}. The similar deviation is also present in the result of GW200105.}
    \label{dipole_dev}
\end{figure}

\begin{figure}
    \centering
    \includegraphics[width=0.5\columnwidth]{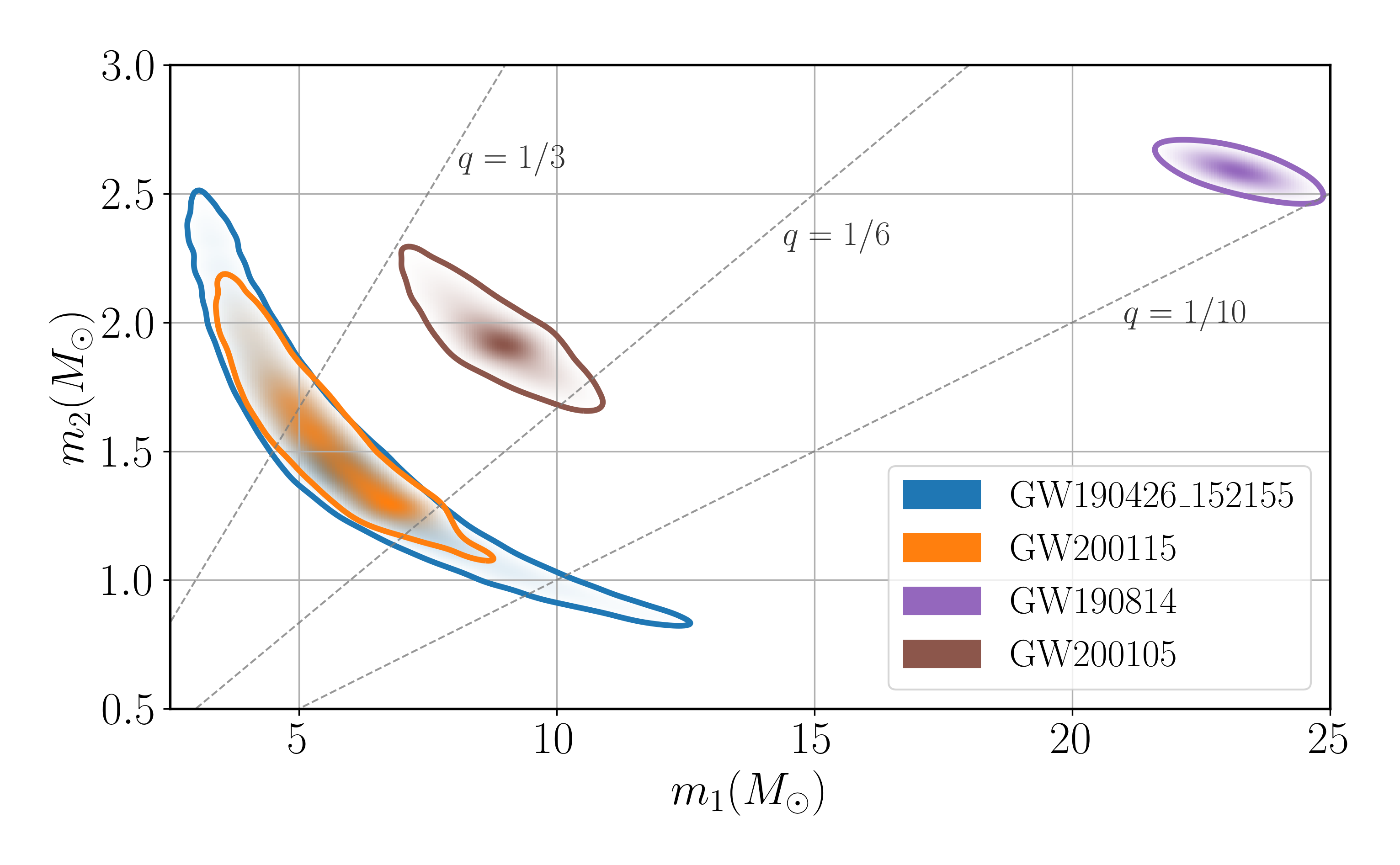}
    \caption{{\bf The component masses of all 4 possible NSBH events so far.} 
    Following  Figure 4 in \citep{Abbott2021a}, we illustrate the component masses of the four possible NSBH events for convenience of reference. The $90\%$ CL regions of the joint posterior distribution for component masses are enclosed by the solid curves, and the shading denotes the posterior probability. The dashed gray lines indicate the constant mass ratio.
    As can be seen in the figure, the GW190814 is the most asymmetric source. Deviations which might be caused by systematic errors of waveform templates, the parameterization method of non-GR modification, or covariances between model parameters, are present in the posteriors of the dipole modification parameter as shown in Figure 19 in \citep{Collaboration2020}, Figure 10 in \citep{Perkins2021} and Figure \ref{dipole_dev} in this paper. 
    Similar deviations are also seen in the case of GW200105, but absent in GW190426\_152155 and GW200115, which is probably due to the more dispersed posteriors and more probability on lower mass ratio.}
    \label{mass_ratio_comparision}
\end{figure}

{ In the main body, we exclude the two events GW190814 and GW200105 due to unphysical deviations. Here, we show the posteriors of the dipole modification parameter for these two events in Figure \ref{dipole_dev} and present more discussion below.

Subdominant spherical harmonic multipoles will become important when the mass ratio of sources is large. There is strong evidence for the presence of higher modes (HMs) in the analysis of GW190814 \citep{Abbott2020a}.
Therefore, following the paper of LVC \citep{Collaboration2020}, we also employ the waveform model incorporating HMs, \texttt{IMRPhenomPv3HM} \citep{Khan2020}. 
The waveform model \texttt{IMRPhenomPv3HM} is based on the model \texttt{IMRPhenomD} \citep{Husa2016,Khan2016} which is employed in the main body, but incorporates the processing due to the in-plane spins and HMs \citep{London2018,Khan2019,Khan2020}.
Same to LVC \citep{Collaboration2020}, we only add the dipole modification on the dominant mode. The non-GR deformation on HMs is gotten by rescaling the modification in the dominant mode according to the method presented in \citep{London2018}. There are no new coefficients introduced.
It is worth to be noted that this method of implementation can possibly be one of the reasons that cause the unphysical deviation.

Using this waveform model, we perform the same Bayesian inference discussed in the main body on the two events GW190814 and GW200105 to constrain the dipole modification parameter $\varphi_{-2}$. The results are shown in Figure \ref{dipole_dev}.
The dashed vertical lines indicate $5\%$ and $95\%$ percentiles for the two events respectively.
It can be seen that the GR value falls in the tails of the posteriors and is excluded from the intervals of $90\%$ CL for these two events. The best fit value of GW190814 deviates from the GR value in the order of $10^{-3}$. While the deviation of GW200105 is slightly reduced. The result of GW190814 presented here is consistent with the result of LVC (as can be seen in Figure 19 of \citep{Collaboration2020}). Similar deviations are also reported in \citep{Perkins2021}.
These results are believed to be not the real deviations from GR. The possible reasons for these deviations might be the systematic errors of the waveform templates or the parameterization method of non-GR modification (which might be inappropriate when HMs are present as pointed out above), and covariances between model parameters \citep{Collaboration2020,Perkins2021}.

We also find the deviations are somehow related to the mass parameters of sources. Referring to Figure 4 in \citep{Abbott2021a}, We also illustrate the component masses of all 4 possible NSBH events so far in Figure \ref{mass_ratio_comparision}. 
The $90\%$ CL regions of the joint posterior distribution for component masses are enclosed by the solid curves, and the shading denotes the posterior probability. The dashed gray lines indicate the constant mass ratio.
The posterior distributions of GW190426\_152155 and GW200115 are almost overlapped. The posteriors of these two events are the most dispersed and have more part in the lower mass ratio. Meanwhile, the deviations on $\varphi_{-2}$ are absent for these two events. The events GW190814 and GW200105 have higher mass ratio, and the magnitude of deviation from GR value is consistent with their mass ratio as can be observed by combining Figures \ref{mass_ratio_comparision} and \ref{dipole_dev}.

For the sources with large mass ratio, the HMs becomes more important which may complicate the analysis. As discussed above, the non-GR modifications on HMs are propagated from the rescalings of the modifications on the dominant mode according to the rules presented in \citep{London2018}. The rescaling rules are verified for the GR part by numerical relativity but are doubtful for the non-GR part. The method of performing the parameterized tests may be inapplicable when HMs are present.
The parameterized tests have not been systematically studied in the parameter space of highly asymmetric sources. More thorough studies are needed to explain these deviations. In this work, we simply exclude the two events GW190814 and GW200105.
}

\bibliography{/Users/hydrogen/Documents/ref/ref}

\bibliographystyle{aasjournal}


\end{document}